\documentclass[12pt,a4paper]{article}
\usepackage{a4wide}
\usepackage{amsmath}
\usepackage{amssymb}
\usepackage{amsfonts}
\usepackage{epsfig}
\usepackage{subfigure}
\usepackage{exscale}
\usepackage{float}
\usepackage{bbm}
\usepackage[numbers,sort&compress]{natbib}
\usepackage{tikz}

\newcommand{\Z}{{\mathbb{Z}}}

\newcommand{\1}{{\mathbbm{1}}}

\makeatletter
\@addtoreset{equation}{section}
\makeatother

\title{Graphical Tensor Product Reduction Scheme for the Lie Algebras
$so(5) = sp(2)$, $su(3)$, and $g(2)$}

\author{N.\ D.\ Vlasii$^{a,b}$, F.\ von R\"utte$^a$, and U.-J.\ Wiese$^a$ \\ \\
{\small $^a$ Albert Einstein Center for Fundamental Physics, 
Institute for Theoretical Physics} \\
{\small Bern University, Sidlerstrasse 5, CH-3012 Bern, Switzerland} \\
{\small $^b$ Bogolyubov Institute for Theoretical Physics, 
National Academy of Sciences of Ukraine} \\
{\small 14-b Metrologichna Str., Kyiv, 03680, Ukraine} \\ \\ \\
We dedicate this paper to the memory of \\ Petro I.\ Holod (1946 - 2014). \\ \\}

\begin{document} 

\maketitle

\begin{abstract} \normalsize

We develop in detail a graphical tensor product reduction scheme, first 
described by Antoine and Speiser, for the simple rank 2 Lie algebras 
$so(5) = sp(2)$, $su(3)$, and $g(2)$. This leads to an efficient practical 
method to reduce tensor products of irreducible representations into sums of 
such representations. For this purpose, the 2-dimensional weight diagram of a 
given representation is placed in a ``landscape'' of irreducible 
representations. We provide both the landscapes and the weight diagrams for a
large number of representations for the three simple rank 2 Lie algebras. We
also apply the algebraic ``girdle'' method, which is much less efficient for 
calculations by hand for moderately large representations. Computer code for 
reducing tensor products, based on the graphical method, has been developed as 
well and is available from the authors upon request.

\end{abstract}

\newpage
 
\section{Introduction}

Besides their fundamental role in mathematics, Lie algebras are of central 
importance in many areas of physics. The Lie algebra $su(2)$ describes rotations
in 3-dimensional coordinate space as well as in the isospin space of nuclear
and particle physics. The Lie algebra $su(3)$ describes the extension of isospin
to the flavor symmetries of up, down, and strange quarks 
\cite{Nee61,Gel62,Zwe64}, and also represents the color degree of freedom by 
which quarks couple to the non-Abelian gluon gauge field 
\cite{Gro73,Pol73,Fri73}. The exceptional Lie algebra $g(2)$ contains $su(3)$ 
as a subalgebra and has been used in early attempts to generalize flavor 
symmetries \cite{Beh62}. The algebra $g(2)$ also plays a role in the context of 
supersymmetry and string theory \cite{Bil02,Egu02,Nis04,HLu04}. The group $G(2)$
has a trivial center, which makes non-Abelian $G(2)$ lattice gauge theories an 
interesting theoretical laboratory for studying confinement and deconfinement
\cite{Hol03,Pep07,Ole08,Wel11,Bru15}. The group $Sp(2) = Spin(5)$, which is the 
universal covering group of $SO(5)$, has the non-trivial center $\Z(2)$. For 
this reason, confinement and deconfinement have also been studied in $Sp(2)$ 
non-Abelian gauge theories \cite{Hol04}. Furthermore, the algebra 
$so(5) = sp(2)$ has been used in condensed matter physics in attempts 
to unify the order parameters of antiferromagnetism and high-temperature 
superconductivity \cite{Zha97,Dem04}. This is certainly just an incomplete list 
of physics applications of the simple rank 2 Lie algebras. In all these and 
many other applications, it is vital to reduce tensor products of irreducible 
representations into sums of such representations.

For the algebras $so(n)$, $sp(n)$, and $su(n)$ there are useful schemes for 
tensor product reduction based on Young tableaux \cite{Ham62,Geo99,Gir82}. While
these develop their full strength for larger values of $n$, they are not the 
most economical schemes for $so(5) = sp(2)$ or $su(3)$. Instead Antoine and 
Speiser have described a graphical method that offers a particularly efficient 
approach to tensor product reduction for the rank 2 Lie algebras 
$so(5) = sp(2)$, $su(3)$, and $g(2)$ \cite{Ant64a,Ant64b}. For this purpose the 
weight diagram of an irreducible representation is placed in a 2-dimensional 
``landscape'' of irreducible representations, centered at its tensor product 
partner. Taking into account the ``parity'' $\pm$ of the various sectors of the 
landscape, the degeneracies of the states in the weight diagram then determine 
the multiplicity with which a given irreducible representation appears in the 
tensor product reduction. An algebraic variant of the graphical method uses 
so-called ``girdles'', which are polynomials associated with each irreducible 
representation \cite{Beh62,Bel69}. By multiplying the girdles and decomposing 
their product into sums of girdles, one then determines the tensor product 
decomposition of the corresponding irreducible representations. Compared to the 
girdle method, the graphical scheme is very intuitive and easy to use by hand 
for moderately large representations, for which the girdle method is already 
very cumbersome. When implemented with automated computer codes, which becomes 
necessary for very large representations, the computational effort of both 
methods is more or less the same. While one of the authors has used the 
graphical Antonie-Speiser scheme for $su(3)$ for several decades, we are 
unaware of places in the literature where the method is worked out in 
sufficient detail to facilitate practical applications. We are also unaware of 
practical applications of the method to the other rank 2 Lie algebras 
$so(5) = sp(2)$ and $g(2)$.

Let us first illustrate the Antoine-Speiser scheme in the simple case of 
$so(3) = su(2) = sp(1)$. The group $SU(2)$ has the center $\Z(2) = \{\pm \1\}$,
consisting of plus and minus the $2 \times 2$ unit-matrix, which obviously
commute with all $SU(2)$ group elements. Correspondingly, there are two
different types of irreducible representations of the $su(2)$ algebra, those
with integer and those with half-integer spin. The integer spin $S$ 
representations $\{2 S + 1\}$ have trivial ``duality'' and contain an odd 
number of $2S + 1$ states, which are distinguished by their spin projection 
$S_3 \in \{-S,-S+1,\dots,S\}$. The half-integer spin representations, on the 
other hand, have non-trivial duality and contain an even number of $2S + 1$ 
states. In contrast to other Lie algebras, all states in an $su(2)$ 
representation are non-degenerate and thus have different values of $S_3$. We
introduce $p = 2 S$ and alternatively denote a representation as 
$(p) = \{2 S + 1\}$. While this alternative notation may seem
unnecessary for $su(2)$, it will be very natural for the higher-rank Lie
algebras. In the interest of a unified notation we therefore introduce $(p)$ 
already for $su(2)$. The dimension of the representation $(p)$ is then given by
\begin{equation}
D(p) = p + 1.
\end{equation}
The weight diagrams of some small representations are illustrated in 
Figs.\ref{su2}a and b. The reduction of the tensor product of two $su(2)$ 
representations with spin $S^a$ and spin $S^b$ results in all representations 
with a total spin $S$ between $|S^a - S^b|$ and $S^a + S^b$, in integer steps, 
i.e.
\begin{equation}
\{2S^a + 1\} \times \{2S^b + 1\} = \{2 |S^a - S^b| + 1\} + \{2 |S^a - S^b| + 3\} 
+ \dots + \{2 (S^a + S^b) + 1\}.
\end{equation}
Indeed the dimensions of the various representations match because one can
easily show that
\begin{equation}
\sum_{S = |S^a - S^b|}^{S = S^a + S^b} (2 S + 1) = (2S^a + 1)(2S^b + 1).
\end{equation}
Tensor product reduction in $su(2)$ is hence so simple that it does not really
require a graphical method. Still, in order to illustrate the Antoine-Speiser
scheme in the simplest case, here we apply it to $su(2)$. Fig.\ref{su2}c shows 
the landscape of $su(2)$ representations with a positive sector on the right 
and a negative sector on the left. The landscape consists of two sublattices 
associated with the integer and half-integer spin representations. In 
Fig.\ref{su2}d the weight diagram of the $S = 2$ representation $\{5\}$ has 
been centered at the position of the $S = \frac{1}{2}$ representation $\{2\}$ 
in the landscape. The five states in $\{5\}$ identify five representations in 
the landscape, which contribute with a positive or negative sign, depending on 
the sector they are in. This leads to the tensor product reduction
\begin{eqnarray}
\label{su2red52}
&&\{5\} \times \{2\} = \{6\} + \{4\} + \{2\} - \{2\} = \{6\} + \{4\} 
\Rightarrow \nonumber \\
&&(4) \times (1) = (5) + (3),
\end{eqnarray}
which indeed corresponds to the total spins $S = \frac{5}{2} = 2 + \frac{1}{2}$ 
and $S = \frac{3}{2} = |2 - \frac{1}{2}|$. Note that we have combined a trivial
duality (integer spin 2) representation with a non-trivial duality 
(half-integer spin $\frac{1}{2}$) representation in the landscape, thus 
resulting in a sum of non-trivial duality (half-integer spin) representations
in the tensor product reduction. The reader may want to copy the weight 
diagrams of Figs.\ref{su2}a and b on a transparency and perform further 
tensor product reductions by superimposing them on a partner representation in 
the landscape of Fig.\ref{su2}c.

Let us also illustrate the girdle method \cite{Beh62,Bel69} in the simple 
$su(2)$ case. First of all, every representation $(p) = \{p + 1\}$ is 
associated with a characteristic Laurent polynomial (containing both positive
and negative powers)
\begin{equation}
\chi(p) = \sum_{n = - p}^p x^n.
\end{equation}
For example, the characteristic Laurent polynomials of the smallest
representations are given by
\begin{eqnarray}
&&\chi(0) = 1, \nonumber \\
&&\chi(1) = x + \frac{1}{x}, \nonumber \\
&&\chi(2) = x^2 + 1 + \frac{1}{x^2}, \nonumber \\
&&\chi(3) = x^3 + x + \frac{1}{x} + \frac{1}{x^3}.
\end{eqnarray}
Tensor product reductions of the representations $(p_1) = \{p_1 + 1\}$ and
$(p_2) = \{p_2 + 1\}$, i.e.
\begin{equation}
(p_1) \times (p_2) = \sum_p n(p_1;p_2;p) (p),
\end{equation}
are then determined from the product of the corresponding characteristic
polynomials
\begin{equation}
\label{redsu2}
\chi(p_1) \chi(p_2) = \sum_p n(p_1;p_2;p) \chi(p).
\end{equation}
Here $n(p_1;p_2;p)$ is the multiplicity with which the representation $(p)$
contributes to the tensor product. For $su(2)$ (but not for the higher-rank
Lie algebras) $n(p_1;p_2;p)$ is limited to 0 or 1.

Interestingly, there is a more efficient way to determine the tensor product 
reduction, in which the characteristic Laurent polynomials are replaced by 
ratios of simpler girdle polynomials 
\begin{equation}
\chi(p) = \frac{\xi(p)}{\xi(0)}, \quad \xi(p) = x^{p+1} - \frac{1}{x^{p+1}}.
\end{equation}
The girdle polynomial is determined by the positions $x = \pm (p+1)$ of the
representation $(p) = \{p+1\}$ in the positive and negative sectors of the
landscape. The girdles associated with the smallest representations are given by
\begin{eqnarray}
&&\xi(0) = x - \frac{1}{x}, \nonumber \\
&&\xi(1) = x^2 - \frac{1}{x^2} = 
\left(x + \frac{1}{x}\right)\left(x - \frac{1}{x}\right) = \chi(1) \xi(0), 
\nonumber \\
&&\xi(2) = x^3 - \frac{1}{x^3} = 
\left(x^2 + 1 + \frac{1}{x^2}\right)\left(x - \frac{1}{x}\right) = 
\chi(2) \xi(0), \nonumber \\
&&\xi(3) = x^4 - \frac{1}{x^4} = 
\left(x^3 + x + \frac{1}{x} + \frac{1}{x^3}\right)
\left(x - \frac{1}{x}\right) = \chi(3) \xi(0).
\end{eqnarray}
This trivially generalizes to all values of $p$ since
\begin{equation}
\xi(p) = x^{p+1} - \frac{1}{x^{p+1}} = 
\sum_{n = - p}^p x^n \left(x - \frac{1}{x}\right) = \chi(p) \xi(0).
\end{equation}
Expressed with girdles, the tensor product decomposition of eq.(\ref{redsu2}) 
then simplifies to
\begin{equation}
\chi(p_1) \xi(p_2) = \sum_p n(p_1;p_2;p) \xi(p).
\end{equation}
For example, the tensor product reduction of eq.(\ref{su2red52}) then takes the
form
\begin{eqnarray}
\chi(4) \xi(1)&=&
\left(x^4 + x^2 + 1 + \frac{1}{x^2} + \frac{1}{x^4}\right) 
\left(x^2 - \frac{1}{x^2}\right) \nonumber \\
&=&x^6 + x^4 + x^2 + 1 + \frac{1}{x^2} - x^2 - 1 - \frac{1}{x^2} - \frac{1}{x^4}
- \frac{1}{x^6} \nonumber \\
&=&\left(x^6 - \frac{1}{x^6}\right) + \left(x^4 - \frac{1}{x^4}\right) =
\xi(5) + \xi(3).
\end{eqnarray}
This again confirms that $\{5\} \times \{2\} = \{6\} + \{4\}$.
 
In this paper, we work out the Antoine-Speiser method in great detail for 
the three simple rank 2 Lie algebras. In particular, we construct the landscapes
for $so(5) = sp(2)$, $su(3)$, and $g(2)$, including relatively large 
representations. We also provide weight diagrams, including the degeneracy
factors of the various states, for a large number of irreducible 
representations. This facilitates the tensor product reduction for a large 
variety of pairs of irreducible representations in a practical and efficient
manner. By copying the weight diagrams on a transparency, the reader can
easily work out rather complicated tensor product reductions by hand. We also
discuss the girdle method, which is much more cumbersome for calculations with 
moderate-size representations by hand, but equivalent to the graphical method
when implemented as an automated computer code. We provide software, available 
from the authors upon request, that implements the graphical Antoine-Speiser
scheme and is applicable to very large irreducible representations. The three 
cases of $so(5) = sp(2)$, $su(3)$, and $g(2)$ are treated in three subsequent 
sections 2, 3, and 4. Section 5 contains our conclusions.

\section{Tensor Product Reduction for $so(5) = sp(2)$}

In this section, we work out the Antoine-Speiser scheme for the algebra $so(5)$
which coincides with the algebra $sp(2)$. After defining these algebras and 
their corresponding Lie groups, we construct the weight diagrams of many 
irreducible representations as well as the corresponding landscape, which is 
2-dimensional because the algebra $so(5) = sp(2)$ has rank 2. We then use the 
method of superimposing a weight diagram on the landscape in order to reduce 
the tensor product of two irreducible representations.

\subsection{The orthogonal group $SO(n)$ and its algebra $so(n)$}

The real-valued $n \times n$ orthogonal matrices $O$ with determinant 1 obey
$O O^T = O^T O = \1$ and form the group $SO(n)$ under matrix multiplication. The
corresponding $so(n)$ algebra consists of the purely imaginary traceless
Hermitean $n \times n$ matrices. There are $n(n-1)/2$ such matrices. The 
algebra $so(4) = su(2) \times su(2)$ is the direct Kronecker product of two
$su(2)$ algebras and thus semi-simple but not simple. 

The group $SO(3)$ has a trivial center, while its universal covering group 
$SU(2)$ has the non-trivial center $\Z(2)$. The universal covering group of 
$SO(n)$ is called $Spin(n)$, such that $Spin(3) = SU(2)$. Similarly, the 
universal covering group of $SO(4)$ is $Spin(4) = SU(2) \times SU(2)$, which 
has the center $\Z(2) \times \Z(2)$. The center of $SO(4)$ itself, on the other 
hand, is just $\Z(2)$ and consists on the $4 \times 4$ unit-matrix $\1$ and 
$- \1$. Since for $n = 5$ the matrix $- \1$ does not have determinant 1, the 
group $SO(5)$ has a trivial center, while its universal covering group 
$Spin(5)$ has the center $\Z(2)$. The group $SO(6)$ has the universal covering
group $Spin(6) = SU(4)$, which has the center $\Z(4)$. At least locally, the 
group manifold of $Spin(n)$ is the product of spheres
\begin{equation}
\label{Spinnmanifold}
Spin(n) = S^1 \times S^2 \dots \times S^{n-1}.
\end{equation}

The $n(n-1)/2$-dimensional adjoint representation of $so(n)$ transforms as an 
anti-symmetric tensor under rotations in $n$ dimensions. Similarly, there is a
representation of dimension $n(n+1)/2 - 1$ that corresponds to a symmetric
traceless tensor. In addition, $so(n)$ has an $n$-dimensional vector 
representation. Since in three dimensions the vector cross product again 
generates a vector (which in this case coincides with an anti-symmetric tensor),
for $so(3)$ the vector representation is equivalent to the adjoint. The $so(n)$ 
algebras also have spinor representations. While $so(3) = su(2)$ only has a 
single 2-dimensional spinor representation $\{2\}$, which corresponds to an 
ordinary spin $\frac{1}{2}$, $so(4) = su(2) \times su(2)$ has two 2-dimensional 
spinor representations, which are both pseudo-real. The algebra $so(5)$ has a 
single 4-dimensional fundamental spinor representation, while $so(6) = su(4)$ 
has two in-equivalent 4-dimensional spinor representations, which correspond to 
the fundamental representation $\{4\}$ of $su(4)$ and its conjugate 
$\{\overline 4\}$. In fact, the $so(n)$ algebras with $n = 6, 10, 14, \dots$ 
are the only ones that have complex representations. 

\subsection{The symplectic group $Sp(n)$ and its algebra $sp(n)$}

The group $Sp(n)$ is a subgroup of $SU(2n)$ leaving the anti-symmetric matrix
\begin{equation}
J = \left(\begin{array}{cc} 0 & \1 \\ - \1 & 0 \end{array}\right) = 
i \sigma^2 \otimes \1,
\end{equation}
invariant, with $\1$ being the $n \times n$ unit-matrix and $\sigma^2$ being the
imaginary Pauli matrix. The elements $U \in SU(2n)$ of the subgroup $Sp(n)$ 
obey the relation
\begin{equation}
\label{pseudoreal}
U^* = J U J^\dagger.
\end{equation}
As a consequence, $U$ and $U^*$ are unitarily equivalent and hence the 
$2n$-dimensional fundamental representation of $Sp(n)$ is pseudo-real. 
It is straightforward to convince oneself that the matrices obeying the
constraint eq.(\ref{pseudoreal}) indeed form a group. The $Sp(n)$ matrices can
be expressed as
\begin{equation}
\label{group}
U =  \left(\begin{array}{cc} W & X \\ - X^* & W^* \end{array}\right), 
\end{equation}
where $W$ and $X$ are complex $n \times n$ matrices. In order for $U$ to still 
belong to $SU(2n)$, the matrices $W$ and $X$ must satisfy $W W^\dagger + 
X X^\dagger = \1$ and $W X^T = X W^T$. The eigenvalues of $U$ occur in
complex conjugate pairs. For center elements, which are multiples of the 
unit-matrix, eq.(\ref{group}) implies $W = W^*$. As a result, the center of 
$Sp(n)$ is $\Z(2)$ for all $n$. Furthermore, $Sp(n)$ is its own universal 
covering group and hence does not give rise to central extensions.

The relation $U = \exp(i H)$, with $H$ being a Hermitean traceless matrix,
together with eq.(\ref{pseudoreal}) implies that the $sp(n)$ generators $H$ 
obey
\begin{equation}
\label{constraint}
H^* = - J H J^\dagger = J H J.
\end{equation}
This relation implies
\begin{equation}
H =  \left(\begin{array}{cc} A & B \\ B^* & - A^* \end{array}\right),
\end{equation}
with $A$ and $B$ being $n \times n$ matrices. The Hermiticity condition $H =
H^\dagger$ leads to $A = A^\dagger$ and $B = B^T$. Since $A$ is Hermitean, $H$ is 
automatically traceless. As a Hermitean $n \times n$ matrix, $A$ has $n^2$ 
degrees of freedom. In addition, the complex symmetric $n \times n$ matrix $B$ 
has $n(n + 1)$ degrees of freedom. Thus $sp(n)$ has the dimension
$n^2 + n(n + 1) = n(2n + 1)$. The algebra $sp(n)$ has $n$ independent diagonal 
generators of its maximal Abelian Cartan subalgebra, such that the rank of 
$sp(n)$ is $n$. 

The algebra $sp(1)$ is equivalent to $so(3) = su(2)$, while $sp(2)$ is 
equivalent to $so(5)$. Since the group $Sp(n)$ has the center $\Z(2)$ while 
$SO(3)$ and $SO(5)$ have a trivial center, the group $Sp(1)$ corresponds to the 
universal covering group $Spin(3) = SU(2)$ of $SO(3)$, and the group $Sp(2)$ is 
the universal covering group $Spin(5)$ of $SO(5)$. Although both $sp(n)$ and 
$so(2n+1)$ have the same number of $n(2n + 1)$ generators, the two algebras
are in-equivalent for $n \geq 3$. Locally, the group manifold of $Sp(n)$ is the 
product of spheres
\begin{equation}
\label{Spnmanifold}
Sp(n) = S^3 \times S^7 \times \dots \times S^{4n-1},
\end{equation}
which implies
\begin{equation}
Sp(1) = S^3 = SU(2), \quad Sp(2) = S^3 \times S^7 = 
S^1 \times S^2 \times S^3 \times S^4 = Spin(5).
\end{equation}
Here we have used the Hopf fibration relations
\begin{equation}
S^3 = S^1 \times S^2, \quad S^7 = S^3 \times S^4.
\end{equation}
On the other hand, since $S^5 \times S^6 \neq S^{11}$ we have
\begin{equation}
Sp(3) = S^3 \times S^7 \times S^{11} =  
S^1 \times S^2 \times S^3 \times S^4 \times S^{11} \neq
S^1 \times S^2 \times S^3 \times S^4 \times S^5 \times S^6 = Spin(7).
\end{equation}

\subsection{Weight diagrams of $so(5) = sp(2)$ representations}

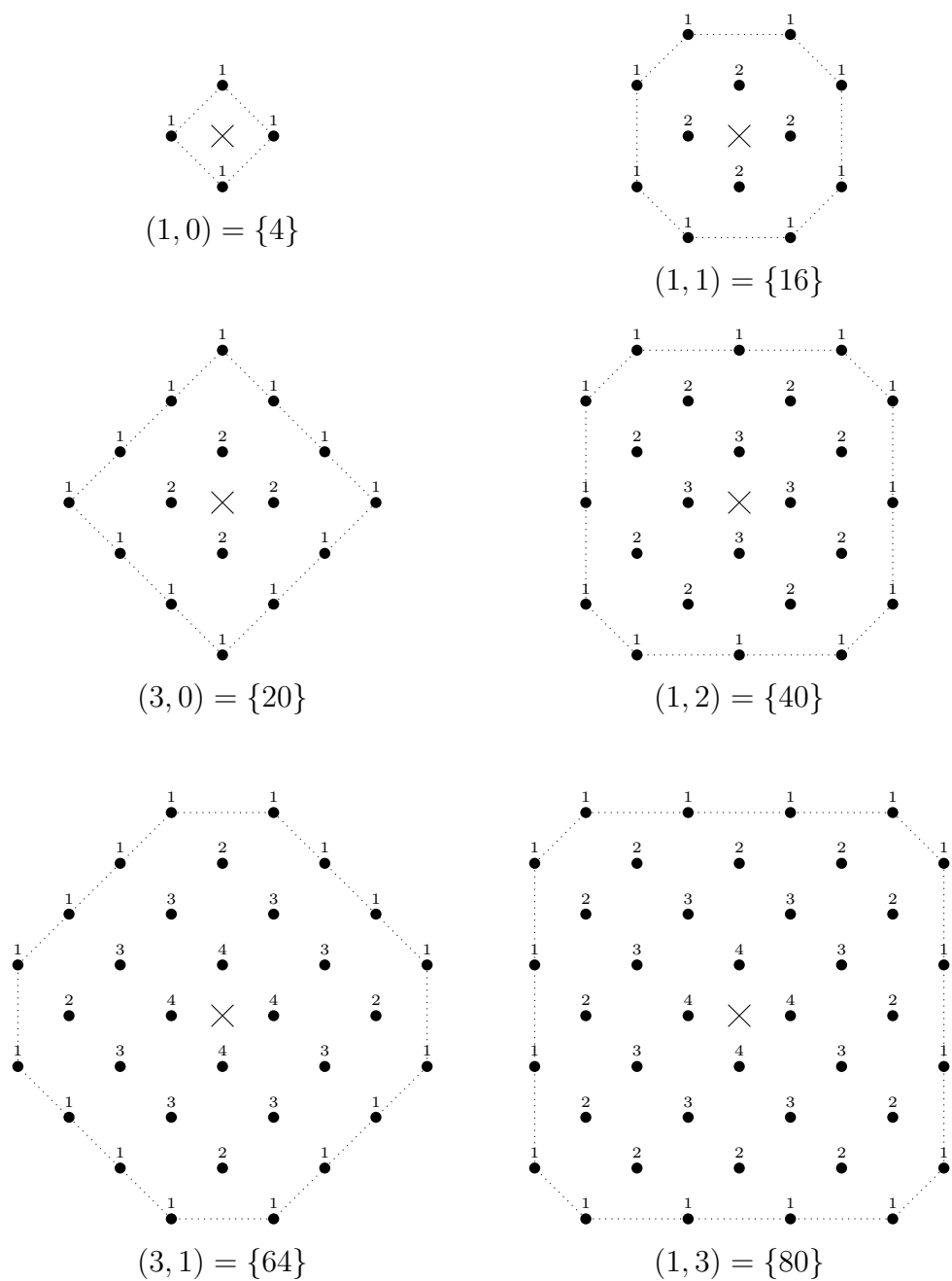
\begin{figure}[tbh]
\begin{center}

\begin{tikzpicture}

\draw ( -4.1500, -0.1500)--( -3.8500,  0.1500);
\draw ( -3.8500, -0.1500)--( -4.1500,  0.1500);
\draw [dotted] ( -3.3072,  0.0000)--( -3.3072,  0.0000);
\draw [dotted] ( -3.3072,  0.0000)--( -4.0000, -0.6928);
\draw [dotted] ( -4.0000, -0.6928)--( -4.0000, -0.6928);
\draw [dotted] ( -4.0000, -0.6928)--( -4.6928,  0.0000);
\draw [dotted] ( -4.6928,  0.0000)--( -4.6928,  0.0000);
\draw [dotted] ( -4.6928,  0.0000)--( -4.0000,  0.6928);
\draw [dotted] ( -4.0000,  0.6928)--( -4.0000,  0.6928);
\draw [dotted] ( -4.0000,  0.6928)--( -3.3072,  0.0000);
\coordinate[label=above:\tiny {$    1$}] (A) at ( -3.3072,  0.0000);
\fill ( -3.3072,  0.0000) circle (0.075);
\coordinate[label=above:\tiny {$    1$}] (A) at ( -4.0000,  0.6928);
\fill ( -4.0000,  0.6928) circle (0.075);
\coordinate[label=above:\tiny {$    1$}] (A) at ( -4.6928,  0.0000);
\fill ( -4.6928,  0.0000) circle (0.075);
\coordinate[label=above:\tiny {$    1$}] (A) at ( -4.0000, -0.6928);
\fill ( -4.0000, -0.6928) circle (0.075);
\coordinate[label=below:{$( 1,0) = \{  4\}$}] (A) at ( -4.0000, -0.9428);

\draw (  2.8500, -0.1500)--(  3.1500,  0.1500);
\draw (  3.1500, -0.1500)--(  2.8500,  0.1500);
\draw [dotted] (  4.3856,  0.6928)--(  4.3856, -0.6928);
\draw [dotted] (  4.3856, -0.6928)--(  3.6928, -1.3856);
\draw [dotted] (  3.6928, -1.3856)--(  2.3072, -1.3856);
\draw [dotted] (  2.3072, -1.3856)--(  1.6144, -0.6928);
\draw [dotted] (  1.6144, -0.6928)--(  1.6144,  0.6928);
\draw [dotted] (  1.6144,  0.6928)--(  2.3072,  1.3856);
\draw [dotted] (  2.3072,  1.3856)--(  3.6928,  1.3856);
\draw [dotted] (  3.6928,  1.3856)--(  4.3856,  0.6928);
\coordinate[label=above:\tiny {$    1$}] (A) at (  4.3856,  0.6928);
\fill (  4.3856,  0.6928) circle (0.075);
\coordinate[label=above:\tiny {$    2$}] (A) at (  3.6928,  0.0000);
\fill (  3.6928,  0.0000) circle (0.075);
\coordinate[label=above:\tiny {$    2$}] (A) at (  3.0000,  0.6928);
\fill (  3.0000,  0.6928) circle (0.075);
\coordinate[label=above:\tiny {$    2$}] (A) at (  2.3072,  0.0000);
\fill (  2.3072,  0.0000) circle (0.075);
\coordinate[label=above:\tiny {$    2$}] (A) at (  3.0000, -0.6928);
\fill (  3.0000, -0.6928) circle (0.075);
\coordinate[label=above:\tiny {$    1$}] (A) at (  3.6928,  1.3856);
\fill (  3.6928,  1.3856) circle (0.075);
\coordinate[label=above:\tiny {$    1$}] (A) at (  2.3072,  1.3856);
\fill (  2.3072,  1.3856) circle (0.075);
\coordinate[label=above:\tiny {$    1$}] (A) at (  1.6144,  0.6928);
\fill (  1.6144,  0.6928) circle (0.075);
\coordinate[label=above:\tiny {$    1$}] (A) at (  1.6144, -0.6928);
\fill (  1.6144, -0.6928) circle (0.075);
\coordinate[label=above:\tiny {$    1$}] (A) at (  2.3072, -1.3856);
\fill (  2.3072, -1.3856) circle (0.075);
\coordinate[label=above:\tiny {$    1$}] (A) at (  3.6928, -1.3856);
\fill (  3.6928, -1.3856) circle (0.075);
\coordinate[label=above:\tiny {$    1$}] (A) at (  4.3856, -0.6928);
\fill (  4.3856, -0.6928) circle (0.075);
\coordinate[label=below:{$( 1,1) = \{ 16\}$}] (A) at (  3.0000, -1.6356);

\draw ( -4.1500, -5.1500)--( -3.8500, -4.8500);
\draw ( -3.8500, -5.1500)--( -4.1500, -4.8500);
\draw [dotted] ( -1.9215, -5.0000)--( -1.9215, -5.0000);
\draw [dotted] ( -1.9215, -5.0000)--( -4.0000, -7.0785);
\draw [dotted] ( -4.0000, -7.0785)--( -4.0000, -7.0785);
\draw [dotted] ( -4.0000, -7.0785)--( -6.0785, -5.0000);
\draw [dotted] ( -6.0785, -5.0000)--( -6.0785, -5.0000);
\draw [dotted] ( -6.0785, -5.0000)--( -4.0000, -2.9215);
\draw [dotted] ( -4.0000, -2.9215)--( -4.0000, -2.9215);
\draw [dotted] ( -4.0000, -2.9215)--( -1.9215, -5.0000);
\coordinate[label=above:\tiny {$    1$}] (A) at ( -1.9215, -5.0000);
\fill ( -1.9215, -5.0000) circle (0.075);
\coordinate[label=above:\tiny {$    1$}] (A) at ( -2.6144, -4.3072);
\fill ( -2.6144, -4.3072) circle (0.075);
\coordinate[label=above:\tiny {$    2$}] (A) at ( -3.3072, -5.0000);
\fill ( -3.3072, -5.0000) circle (0.075);
\coordinate[label=above:\tiny {$    2$}] (A) at ( -4.0000, -4.3072);
\fill ( -4.0000, -4.3072) circle (0.075);
\coordinate[label=above:\tiny {$    2$}] (A) at ( -4.6928, -5.0000);
\fill ( -4.6928, -5.0000) circle (0.075);
\coordinate[label=above:\tiny {$    2$}] (A) at ( -4.0000, -5.6928);
\fill ( -4.0000, -5.6928) circle (0.075);
\coordinate[label=above:\tiny {$    1$}] (A) at ( -3.3072, -3.6144);
\fill ( -3.3072, -3.6144) circle (0.075);
\coordinate[label=above:\tiny {$    1$}] (A) at ( -4.6928, -3.6144);
\fill ( -4.6928, -3.6144) circle (0.075);
\coordinate[label=above:\tiny {$    1$}] (A) at ( -5.3856, -4.3072);
\fill ( -5.3856, -4.3072) circle (0.075);
\coordinate[label=above:\tiny {$    1$}] (A) at ( -5.3856, -5.6928);
\fill ( -5.3856, -5.6928) circle (0.075);
\coordinate[label=above:\tiny {$    1$}] (A) at ( -4.6928, -6.3856);
\fill ( -4.6928, -6.3856) circle (0.075);
\coordinate[label=above:\tiny {$    1$}] (A) at ( -3.3072, -6.3856);
\fill ( -3.3072, -6.3856) circle (0.075);
\coordinate[label=above:\tiny {$    1$}] (A) at ( -2.6144, -5.6928);
\fill ( -2.6144, -5.6928) circle (0.075);
\coordinate[label=above:\tiny {$    1$}] (A) at ( -4.0000, -2.9215);
\fill ( -4.0000, -2.9215) circle (0.075);
\coordinate[label=above:\tiny {$    1$}] (A) at ( -6.0785, -5.0000);
\fill ( -6.0785, -5.0000) circle (0.075);
\coordinate[label=above:\tiny {$    1$}] (A) at ( -4.0000, -7.0785);
\fill ( -4.0000, -7.0785) circle (0.075);
\coordinate[label=below:{$( 3,0) = \{ 20\}$}] (A) at ( -4.0000, -7.3285);

\draw (  2.8500, -5.1500)--(  3.1500, -4.8500);
\draw (  3.1500, -5.1500)--(  2.8500, -4.8500);
\draw [dotted] (  5.0785, -3.6144)--(  5.0785, -6.3856);
\draw [dotted] (  5.0785, -6.3856)--(  4.3856, -7.0785);
\draw [dotted] (  4.3856, -7.0785)--(  1.6144, -7.0785);
\draw [dotted] (  1.6144, -7.0785)--(  0.9215, -6.3856);
\draw [dotted] (  0.9215, -6.3856)--(  0.9215, -3.6144);
\draw [dotted] (  0.9215, -3.6144)--(  1.6144, -2.9215);
\draw [dotted] (  1.6144, -2.9215)--(  4.3856, -2.9215);
\draw [dotted] (  4.3856, -2.9215)--(  5.0785, -3.6144);
\coordinate[label=above:\tiny {$    1$}] (A) at (  5.0785, -3.6144);
\fill (  5.0785, -3.6144) circle (0.075);
\coordinate[label=above:\tiny {$    1$}] (A) at (  5.0785, -5.0000);
\fill (  5.0785, -5.0000) circle (0.075);
\coordinate[label=above:\tiny {$    2$}] (A) at (  4.3856, -4.3072);
\fill (  4.3856, -4.3072) circle (0.075);
\coordinate[label=above:\tiny {$    3$}] (A) at (  3.6928, -5.0000);
\fill (  3.6928, -5.0000) circle (0.075);
\coordinate[label=above:\tiny {$    3$}] (A) at (  3.0000, -4.3072);
\fill (  3.0000, -4.3072) circle (0.075);
\coordinate[label=above:\tiny {$    3$}] (A) at (  2.3072, -5.0000);
\fill (  2.3072, -5.0000) circle (0.075);
\coordinate[label=above:\tiny {$    3$}] (A) at (  3.0000, -5.6928);
\fill (  3.0000, -5.6928) circle (0.075);
\coordinate[label=above:\tiny {$    2$}] (A) at (  3.6928, -3.6144);
\fill (  3.6928, -3.6144) circle (0.075);
\coordinate[label=above:\tiny {$    2$}] (A) at (  2.3072, -3.6144);
\fill (  2.3072, -3.6144) circle (0.075);
\coordinate[label=above:\tiny {$    2$}] (A) at (  1.6144, -4.3072);
\fill (  1.6144, -4.3072) circle (0.075);
\coordinate[label=above:\tiny {$    2$}] (A) at (  1.6144, -5.6928);
\fill (  1.6144, -5.6928) circle (0.075);
\coordinate[label=above:\tiny {$    2$}] (A) at (  2.3072, -6.3856);
\fill (  2.3072, -6.3856) circle (0.075);
\coordinate[label=above:\tiny {$    2$}] (A) at (  3.6928, -6.3856);
\fill (  3.6928, -6.3856) circle (0.075);
\coordinate[label=above:\tiny {$    2$}] (A) at (  4.3856, -5.6928);
\fill (  4.3856, -5.6928) circle (0.075);
\coordinate[label=above:\tiny {$    1$}] (A) at (  3.0000, -2.9215);
\fill (  3.0000, -2.9215) circle (0.075);
\coordinate[label=above:\tiny {$    1$}] (A) at (  0.9215, -5.0000);
\fill (  0.9215, -5.0000) circle (0.075);
\coordinate[label=above:\tiny {$    1$}] (A) at (  3.0000, -7.0785);
\fill (  3.0000, -7.0785) circle (0.075);
\coordinate[label=above:\tiny {$    1$}] (A) at (  4.3856, -2.9215);
\fill (  4.3856, -2.9215) circle (0.075);
\coordinate[label=above:\tiny {$    1$}] (A) at (  1.6144, -2.9215);
\fill (  1.6144, -2.9215) circle (0.075);
\coordinate[label=above:\tiny {$    1$}] (A) at (  0.9215, -3.6144);
\fill (  0.9215, -3.6144) circle (0.075);
\coordinate[label=above:\tiny {$    1$}] (A) at (  0.9215, -6.3856);
\fill (  0.9215, -6.3856) circle (0.075);
\coordinate[label=above:\tiny {$    1$}] (A) at (  1.6144, -7.0785);
\fill (  1.6144, -7.0785) circle (0.075);
\coordinate[label=above:\tiny {$    1$}] (A) at (  4.3856, -7.0785);
\fill (  4.3856, -7.0785) circle (0.075);
\coordinate[label=above:\tiny {$    1$}] (A) at (  5.0785, -6.3856);
\fill (  5.0785, -6.3856) circle (0.075);
\coordinate[label=below:{$( 1,2) = \{ 40\}$}] (A) at (  3.0000, -7.3285);

\draw ( -4.1500,-12.1500)--( -3.8500,-11.8500);
\draw ( -3.8500,-12.1500)--( -4.1500,-11.8500);
\draw [dotted] ( -1.2287,-11.3072)--( -1.2287,-12.6928);
\draw [dotted] ( -1.2287,-12.6928)--( -3.3072,-14.7713);
\draw [dotted] ( -3.3072,-14.7713)--( -4.6928,-14.7713);
\draw [dotted] ( -4.6928,-14.7713)--( -6.7713,-12.6928);
\draw [dotted] ( -6.7713,-12.6928)--( -6.7713,-11.3072);
\draw [dotted] ( -6.7713,-11.3072)--( -4.6928, -9.2287);
\draw [dotted] ( -4.6928, -9.2287)--( -3.3072, -9.2287);
\draw [dotted] ( -3.3072, -9.2287)--( -1.2287,-11.3072);
\coordinate[label=above:\tiny {$    1$}] (A) at ( -1.2287,-11.3072);
\fill ( -1.2287,-11.3072) circle (0.075);
\coordinate[label=above:\tiny {$    1$}] (A) at ( -1.9215,-10.6144);
\fill ( -1.9215,-10.6144) circle (0.075);
\coordinate[label=above:\tiny {$    2$}] (A) at ( -1.9215,-12.0000);
\fill ( -1.9215,-12.0000) circle (0.075);
\coordinate[label=above:\tiny {$    3$}] (A) at ( -2.6144,-11.3072);
\fill ( -2.6144,-11.3072) circle (0.075);
\coordinate[label=above:\tiny {$    4$}] (A) at ( -3.3072,-12.0000);
\fill ( -3.3072,-12.0000) circle (0.075);
\coordinate[label=above:\tiny {$    4$}] (A) at ( -4.0000,-11.3072);
\fill ( -4.0000,-11.3072) circle (0.075);
\coordinate[label=above:\tiny {$    4$}] (A) at ( -4.6928,-12.0000);
\fill ( -4.6928,-12.0000) circle (0.075);
\coordinate[label=above:\tiny {$    4$}] (A) at ( -4.0000,-12.6928);
\fill ( -4.0000,-12.6928) circle (0.075);
\coordinate[label=above:\tiny {$    3$}] (A) at ( -3.3072,-10.6144);
\fill ( -3.3072,-10.6144) circle (0.075);
\coordinate[label=above:\tiny {$    3$}] (A) at ( -4.6928,-10.6144);
\fill ( -4.6928,-10.6144) circle (0.075);
\coordinate[label=above:\tiny {$    3$}] (A) at ( -5.3856,-11.3072);
\fill ( -5.3856,-11.3072) circle (0.075);
\coordinate[label=above:\tiny {$    3$}] (A) at ( -5.3856,-12.6928);
\fill ( -5.3856,-12.6928) circle (0.075);
\coordinate[label=above:\tiny {$    3$}] (A) at ( -4.6928,-13.3856);
\fill ( -4.6928,-13.3856) circle (0.075);
\coordinate[label=above:\tiny {$    3$}] (A) at ( -3.3072,-13.3856);
\fill ( -3.3072,-13.3856) circle (0.075);
\coordinate[label=above:\tiny {$    3$}] (A) at ( -2.6144,-12.6928);
\fill ( -2.6144,-12.6928) circle (0.075);
\coordinate[label=above:\tiny {$    2$}] (A) at ( -4.0000, -9.9215);
\fill ( -4.0000, -9.9215) circle (0.075);
\coordinate[label=above:\tiny {$    2$}] (A) at ( -6.0785,-12.0000);
\fill ( -6.0785,-12.0000) circle (0.075);
\coordinate[label=above:\tiny {$    2$}] (A) at ( -4.0000,-14.0785);
\fill ( -4.0000,-14.0785) circle (0.075);
\coordinate[label=above:\tiny {$    1$}] (A) at ( -2.6144, -9.9215);
\fill ( -2.6144, -9.9215) circle (0.075);
\coordinate[label=above:\tiny {$    1$}] (A) at ( -5.3856, -9.9215);
\fill ( -5.3856, -9.9215) circle (0.075);
\coordinate[label=above:\tiny {$    1$}] (A) at ( -6.0785,-10.6144);
\fill ( -6.0785,-10.6144) circle (0.075);
\coordinate[label=above:\tiny {$    1$}] (A) at ( -6.0785,-13.3856);
\fill ( -6.0785,-13.3856) circle (0.075);
\coordinate[label=above:\tiny {$    1$}] (A) at ( -5.3856,-14.0785);
\fill ( -5.3856,-14.0785) circle (0.075);
\coordinate[label=above:\tiny {$    1$}] (A) at ( -2.6144,-14.0785);
\fill ( -2.6144,-14.0785) circle (0.075);
\coordinate[label=above:\tiny {$    1$}] (A) at ( -1.9215,-13.3856);
\fill ( -1.9215,-13.3856) circle (0.075);
\coordinate[label=above:\tiny {$    1$}] (A) at ( -3.3072, -9.2287);
\fill ( -3.3072, -9.2287) circle (0.075);
\coordinate[label=above:\tiny {$    1$}] (A) at ( -4.6928, -9.2287);
\fill ( -4.6928, -9.2287) circle (0.075);
\coordinate[label=above:\tiny {$    1$}] (A) at ( -6.7713,-11.3072);
\fill ( -6.7713,-11.3072) circle (0.075);
\coordinate[label=above:\tiny {$    1$}] (A) at ( -6.7713,-12.6928);
\fill ( -6.7713,-12.6928) circle (0.075);
\coordinate[label=above:\tiny {$    1$}] (A) at ( -4.6928,-14.7713);
\fill ( -4.6928,-14.7713) circle (0.075);
\coordinate[label=above:\tiny {$    1$}] (A) at ( -3.3072,-14.7713);
\fill ( -3.3072,-14.7713) circle (0.075);
\coordinate[label=above:\tiny {$    1$}] (A) at ( -1.2287,-12.6928);
\fill ( -1.2287,-12.6928) circle (0.075);
\coordinate[label=below:{$( 3,1) = \{ 64\}$}] (A) at ( -4.0000,-15.0213);

\draw (  2.8500,-12.1500)--(  3.1500,-11.8500);
\draw (  3.1500,-12.1500)--(  2.8500,-11.8500);
\draw [dotted] (  5.7713, -9.9215)--(  5.7713,-14.0785);
\draw [dotted] (  5.7713,-14.0785)--(  5.0785,-14.7713);
\draw [dotted] (  5.0785,-14.7713)--(  0.9215,-14.7713);
\draw [dotted] (  0.9215,-14.7713)--(  0.2287,-14.0785);
\draw [dotted] (  0.2287,-14.0785)--(  0.2287, -9.9215);
\draw [dotted] (  0.2287, -9.9215)--(  0.9215, -9.2287);
\draw [dotted] (  0.9215, -9.2287)--(  5.0785, -9.2287);
\draw [dotted] (  5.0785, -9.2287)--(  5.7713, -9.9215);
\coordinate[label=above:\tiny {$    1$}] (A) at (  5.7713, -9.9215);
\fill (  5.7713, -9.9215) circle (0.075);
\coordinate[label=above:\tiny {$    1$}] (A) at (  5.7713,-11.3072);
\fill (  5.7713,-11.3072) circle (0.075);
\coordinate[label=above:\tiny {$    2$}] (A) at (  5.0785,-10.6144);
\fill (  5.0785,-10.6144) circle (0.075);
\coordinate[label=above:\tiny {$    2$}] (A) at (  5.0785,-12.0000);
\fill (  5.0785,-12.0000) circle (0.075);
\coordinate[label=above:\tiny {$    3$}] (A) at (  4.3856,-11.3072);
\fill (  4.3856,-11.3072) circle (0.075);
\coordinate[label=above:\tiny {$    4$}] (A) at (  3.6928,-12.0000);
\fill (  3.6928,-12.0000) circle (0.075);
\coordinate[label=above:\tiny {$    4$}] (A) at (  3.0000,-11.3072);
\fill (  3.0000,-11.3072) circle (0.075);
\coordinate[label=above:\tiny {$    4$}] (A) at (  2.3072,-12.0000);
\fill (  2.3072,-12.0000) circle (0.075);
\coordinate[label=above:\tiny {$    4$}] (A) at (  3.0000,-12.6928);
\fill (  3.0000,-12.6928) circle (0.075);
\coordinate[label=above:\tiny {$    3$}] (A) at (  3.6928,-10.6144);
\fill (  3.6928,-10.6144) circle (0.075);
\coordinate[label=above:\tiny {$    3$}] (A) at (  2.3072,-10.6144);
\fill (  2.3072,-10.6144) circle (0.075);
\coordinate[label=above:\tiny {$    3$}] (A) at (  1.6144,-11.3072);
\fill (  1.6144,-11.3072) circle (0.075);
\coordinate[label=above:\tiny {$    3$}] (A) at (  1.6144,-12.6928);
\fill (  1.6144,-12.6928) circle (0.075);
\coordinate[label=above:\tiny {$    3$}] (A) at (  2.3072,-13.3856);
\fill (  2.3072,-13.3856) circle (0.075);
\coordinate[label=above:\tiny {$    3$}] (A) at (  3.6928,-13.3856);
\fill (  3.6928,-13.3856) circle (0.075);
\coordinate[label=above:\tiny {$    3$}] (A) at (  4.3856,-12.6928);
\fill (  4.3856,-12.6928) circle (0.075);
\coordinate[label=above:\tiny {$    2$}] (A) at (  3.0000, -9.9215);
\fill (  3.0000, -9.9215) circle (0.075);
\coordinate[label=above:\tiny {$    2$}] (A) at (  0.9215,-12.0000);
\fill (  0.9215,-12.0000) circle (0.075);
\coordinate[label=above:\tiny {$    2$}] (A) at (  3.0000,-14.0785);
\fill (  3.0000,-14.0785) circle (0.075);
\coordinate[label=above:\tiny {$    2$}] (A) at (  4.3856, -9.9215);
\fill (  4.3856, -9.9215) circle (0.075);
\coordinate[label=above:\tiny {$    2$}] (A) at (  1.6144, -9.9215);
\fill (  1.6144, -9.9215) circle (0.075);
\coordinate[label=above:\tiny {$    2$}] (A) at (  0.9215,-10.6144);
\fill (  0.9215,-10.6144) circle (0.075);
\coordinate[label=above:\tiny {$    2$}] (A) at (  0.9215,-13.3856);
\fill (  0.9215,-13.3856) circle (0.075);
\coordinate[label=above:\tiny {$    2$}] (A) at (  1.6144,-14.0785);
\fill (  1.6144,-14.0785) circle (0.075);
\coordinate[label=above:\tiny {$    2$}] (A) at (  4.3856,-14.0785);
\fill (  4.3856,-14.0785) circle (0.075);
\coordinate[label=above:\tiny {$    2$}] (A) at (  5.0785,-13.3856);
\fill (  5.0785,-13.3856) circle (0.075);
\coordinate[label=above:\tiny {$    1$}] (A) at (  3.6928, -9.2287);
\fill (  3.6928, -9.2287) circle (0.075);
\coordinate[label=above:\tiny {$    1$}] (A) at (  2.3072, -9.2287);
\fill (  2.3072, -9.2287) circle (0.075);
\coordinate[label=above:\tiny {$    1$}] (A) at (  0.2287,-11.3072);
\fill (  0.2287,-11.3072) circle (0.075);
\coordinate[label=above:\tiny {$    1$}] (A) at (  0.2287,-12.6928);
\fill (  0.2287,-12.6928) circle (0.075);
\coordinate[label=above:\tiny {$    1$}] (A) at (  2.3072,-14.7713);
\fill (  2.3072,-14.7713) circle (0.075);
\coordinate[label=above:\tiny {$    1$}] (A) at (  3.6928,-14.7713);
\fill (  3.6928,-14.7713) circle (0.075);
\coordinate[label=above:\tiny {$    1$}] (A) at (  5.7713,-12.6928);
\fill (  5.7713,-12.6928) circle (0.075);
\coordinate[label=above:\tiny {$    1$}] (A) at (  5.0785, -9.2287);
\fill (  5.0785, -9.2287) circle (0.075);
\coordinate[label=above:\tiny {$    1$}] (A) at (  0.9215, -9.2287);
\fill (  0.9215, -9.2287) circle (0.075);
\coordinate[label=above:\tiny {$    1$}] (A) at (  0.2287, -9.9215);
\fill (  0.2287, -9.9215) circle (0.075);
\coordinate[label=above:\tiny {$    1$}] (A) at (  0.2287,-14.0785);
\fill (  0.2287,-14.0785) circle (0.075);
\coordinate[label=above:\tiny {$    1$}] (A) at (  0.9215,-14.7713);
\fill (  0.9215,-14.7713) circle (0.075);
\coordinate[label=above:\tiny {$    1$}] (A) at (  5.0785,-14.7713);
\fill (  5.0785,-14.7713) circle (0.075);
\coordinate[label=above:\tiny {$    1$}] (A) at (  5.7713,-14.0785);
\fill (  5.7713,-14.0785) circle (0.075);
\coordinate[label=below:{$( 1,3) = \{ 80\}$}] (A) at (  3.0000,-15.0213);

\end{tikzpicture}

\end{center}
\caption{\it The weight diagrams of several $so(5) = sp(2)$ representations of
non-trivial duality. The axes are labeled by the eigenvalues of the commuting 
generators $T_3^1$ and $T_3^2$ of the subalgebra $so(4) = su(2) \times su(2)$.
Note that for representations of non-trivial duality the origin is not 
occupied by a state in the weight diagram. The superscripts denote the
degeneracies of the various states. A representation $(p,q)$ (which we 
alternatively denote as $\{D(p,q)\}$) is uniquely characterized by the side 
lengths $p$ (along a diagonal) and $q$ (along the Cartesian axes) of its 
octagon-shaped weight diagram, that determine its dimension $D(p,q)$.}
\label{sp2weighta}
\end{figure}

\begin{figure}[tbh]
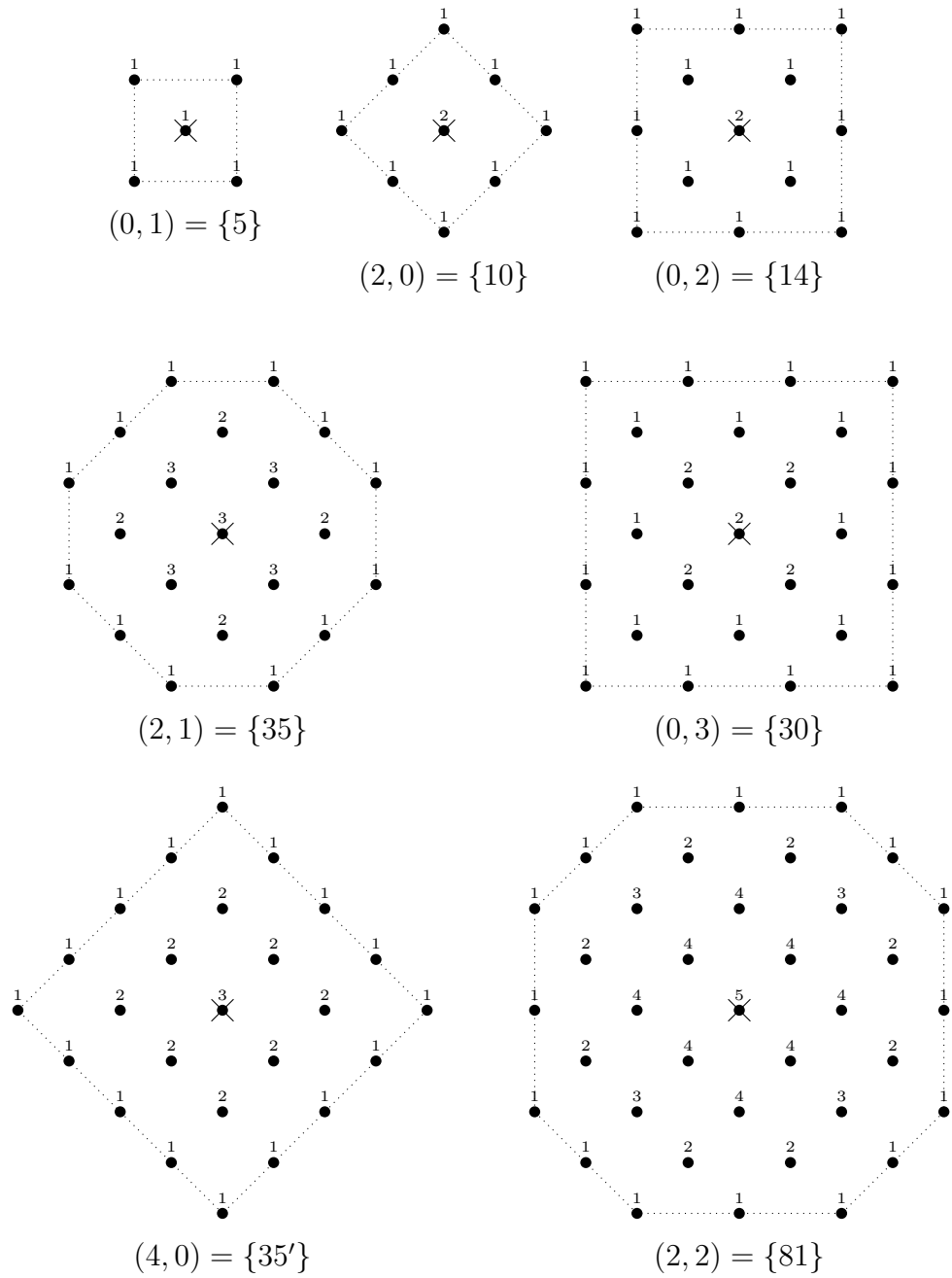

\begin{center}



\end{center}
\caption{\it The weight diagrams of several $so(5) = sp(2)$ representations of
trivial duality. In this case, the origin is always occupied by a state in the 
weight diagram.}
\label{sp2weightb}
\end{figure}

\begin{figure}[tbh]
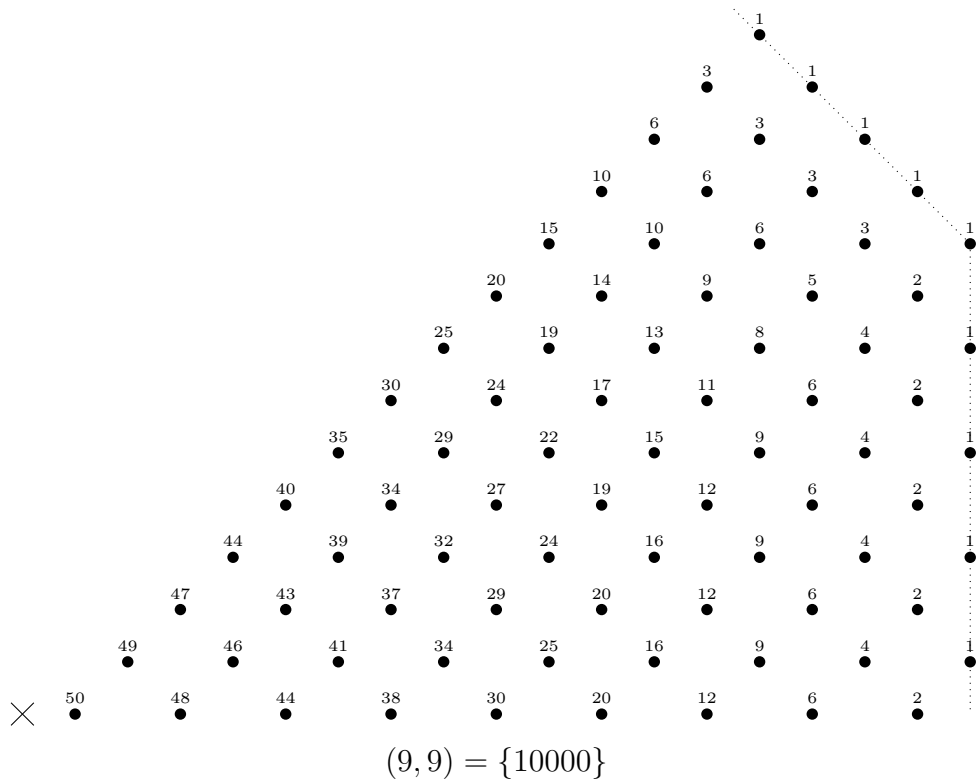

\begin{center}

%

\end{center}
\caption{\it One of eight segments of the weight diagram of the $so(5) = sp(2)$ 
representation $(9,9) = \{10000\}$.}
\label{sp2WD10000}
\end{figure}

Since $so(5) = sp(2)$ has rank 2, the weight diagrams of the corresponding 
irreducible representations can be drawn in a 2-dimensional plane. The 
eigenvalues of the commuting generators $T_3^1$ and $T_3^2$ of the subalgebra 
$so(4) = su(2) \times su(2)$ can be used to characterize the states of an
irreducible representation. In contrast to $su(2)$, some states in an 
irreducible $so(5) = sp(2)$ representation may be degenerate, i.e.\ they may 
have the same eigenvalues of $T_3^1$ and $T_3^2$. Just as for $su(2)$, since the
center of $Spin(5) = Sp(2)$ is $\Z(2)$, there are two classes of 
$so(5) = sp(2)$ representations, one of trivial and one of non-trivial duality.
The weight diagrams of several irreducible representations of non-trivial 
duality, including the fundamental representation $\{4\}$, are illustrated in
Fig.\ref{sp2weighta}. In this case, the origin is not occupied by a state in the
weight diagrams. Several representations of trivial duality, including the 
$so(5)$ vector representation $\{5\}$ and the adjoint representation $\{10\}$, 
are depicted in Fig.\ref{sp2weightb}. For representations of trivial duality,
the origin is always occupied by a state in the weight diagrams. The weight
diagram of a general representation has the shape of an octagon, which is
characterized by its side lengths $q$ along the Cartesian axes, and 
$p$ along the diagonals. For representations with trivial duality $p$ is even, 
while for representations with non-trivial duality $p$ is odd. As we will 
discuss later, the degeneracies of the various states in the weight diagram of 
a representation $(p,q)$ can be determined recursively by applying the 
Antoine-Speiser scheme to its tensor product with the trivial representation 
$(0,0) = \{1\}$. The dimension of a representation, i.e.\ its total number of 
states, is determined by $p$ and $q$ and is given by
\begin{equation}
\label{Dsp2}
D(p,q) = \frac{1}{6}(p + 1)(q + 1)(p + q + 2)(p + 2q + 3).
\end{equation}
An irreducible representation $(p,q)$ is completely characterized by the side
lengths $p$ and $q$ of its octagon-shaped weight diagram. However, it is not
uniquely determined by its dimension $D(p,q)$. For example, 
$D(2,1) = D(4,0) = 35$ and $D(0,6) = D(1,4) = D(3,2) = 140$. Still, we 
alternatively denote a representation $(p,q)$ by $\{D(p,q)\}$. In order to 
distinguish the ambiguous cases, we denote $(2,1) = \{35\}$, $(4,0) = \{35'\}$,
$(3,2) = \{140\}$, $(1,4) = \{140'\}$, and $(0,6) = \{140''\}$.  The 
degeneracies for $D(p,q) \leq 10^4$ are listed in Table \ref{tablesp2}. For
$D(p,q) \leq 10^8$ one encounters degeneracies up to $g = 4$, for example,
\begin{equation}
D(0,54) = D(2,36) = D(3,32) = D(29,6) = 56980.
\end{equation}
As a side remark, we like to mention that $(9,9) = \{10000\}$. The weight 
diagram of this representation is illustrated in Fig.\ref{sp2WD10000}.

\begin{table}[tbh]
\begin{center}
\begin{tabular}{|c|c|c|c|c|}
\hline 
$D(p,q)$ & $g$ & $(p_1,q_1)$ & $(p_2,q_2)$ & $(p_3,q_3)$ \\
\hline
\hline
35       &  2  & $(2,1)$     & $(4,0)$    &             \\
\hline
140      &  3  & $(0,6)$     & $(1,4)$    & $(3,2)$     \\
\hline
220      &  2  & $(4,2)$     & $(9,0)$    &             \\
\hline
455      &  2  & $(6,2)$     & $(12,0)$   &             \\
\hline
560      &  3  & $(5,3)$     & $(9,1)$    & $(13,0)$    \\
\hline
880      &  2  & $(1,9)$     & $(5,4)$    &             \\
\hline
1330     &  2  & $(4,6)$     & $(18,0)$   &             \\
\hline
1820     &  3  & $(1,12)$    & $(4,7)$    & $(5,6)$     \\
\hline
2240     &  3  & $(1,13)$    & $(3,9)$    & $(7,5)$     \\
\hline
2835     &  2  & $(8,5)$     & $(14,2)$   &             \\
\hline
3080     &  2  & $(7,6)$     & $(10,4)$   &             \\
\hline
3520     &  2  & $(9,5)$     & $(19,1)$   &             \\
\hline
5320     &  3  & $(1,18)$    & $(2,15)$   & $(13,4)$    \\
\hline
7280     &  3  & $(13,5)$    & $(15,4)$   & $(25,1)$    \\
\hline
8960     &  3  & $(11,7)$    & $(19,3)$   & $(27,1)$    \\
\hline
\end{tabular}
\end{center}
\caption{\it Degeneracies $g \ge 2$ of the dimensions $D(p,q) \leq 10^4$ of 
$so(5) = sp(2)$ representations $(p,q)$. The degeneracy factor $g$ counts in
how many ways $D(p,q)$ can be realized by pairs $(p,q)$.}
\label{tablesp2}
\end{table}

The degeneracies of the various states in a general weight diagram do not 
follow any obvious pattern. However, as one sees in Figs.\ref{sp2weighta} and 
\ref{sp2weightb}, the degeneracies of the states in the diamond-shaped 
weight diagrams of the representation $(p,0)$ follow a shell structure. The 
states in the outer shell at the edge of the weight diagram are not degenerate. 
The states in the next inner shell are two-fold degenerate. As one moves on to 
further interior shells, the degeneracy increases by one. This behavior is
consistent with eq.(\ref{Dsp2}) for $D(p,0)$ which obeys
\begin{eqnarray}
D(p,0)&=&\frac{1}{6} (p + 1)(p + 2)(p + 3) =
\frac{1}{6} (p - 1)p(p + 1) + (p + 1)^2 \nonumber \\
&=&D(p - 2,0) + A(p,0), \qquad p \geq 2.
\end{eqnarray}
Here 
\begin{equation}
A(p,q) = p^2 + 4 p q + 2 q^2 + 2 p + 2 q + 1
\end{equation}
is the ``area'' of the weight diagram of the representation $(p,q)$, i.e.\ the 
number of points in it, not counting their degeneracies. By subtracting 
\begin{equation}
A(p,0) = (p + 1)^2
\end{equation}
from $D(p,0)$, as long as $p \geq 2$, one removes the outer shell of 
the weight diagram of the representation $(p,0)$ and one reduces the 
degeneracies of all other states by 1, thus arriving at the representation 
$(p - 2,0)$. The same shell structure exists for the representations $(1,q)$
for which
\begin{eqnarray}
D(1,q)&=&\frac{2}{3} (q + 1)(q + 2)(q + 3) =
\frac{2}{3} q(q + 1)(q + 2) + 2(q + 1)(q + 2) \nonumber \\
&=&D(1,q - 1) + A(1,q), \qquad q \geq 1,
\end{eqnarray}
where the area of the weight diagram of the representation $(1,q)$ is given by
\begin{equation}
A(1,q) = 2(q + 1)(q + 2).
\end{equation}
Finally, a double shell structure, with two subsequent shells having the same
degeneracy, is observed for the representations $(0,q)$, which have a 
square-shaped weight diagram. For them
\begin{eqnarray}
D(0,q)&=&\frac{1}{6} (q + 1)(q + 2)(2 q + 3) =
\frac{1}{6} (q - 1)q(2 q - 1) + (q + 1)^2 + q^2 \nonumber \\
&=&D(0,q - 2) + A(0,q), \qquad q \geq 1,
\end{eqnarray}
where the area of the weight diagram of the representation $(0,q)$ is given by
\begin{equation}
A(0,q) = (q + 1)^2 + q^2.
\end{equation}

\subsection{Landscape of $so(5) = sp(2)$ representations}

As discussed by Antoine and Speiser \cite{Ant64a,Ant64b}, the representations 
of a rank 2 Lie algebra can be positioned in a 2-dimensional plane, which we 
denote as a landscape. The landscape of $so(5) = sp(2)$ representations is 
depicted in Fig.\ref{sp2landscape}. The representations of trivial and 
non-trivial duality are associated with the points of the odd and even 
sublattices of a square lattice, respectively. 

\begin{figure}[tbp]
\begin{center}
\includegraphics[width=\textwidth]{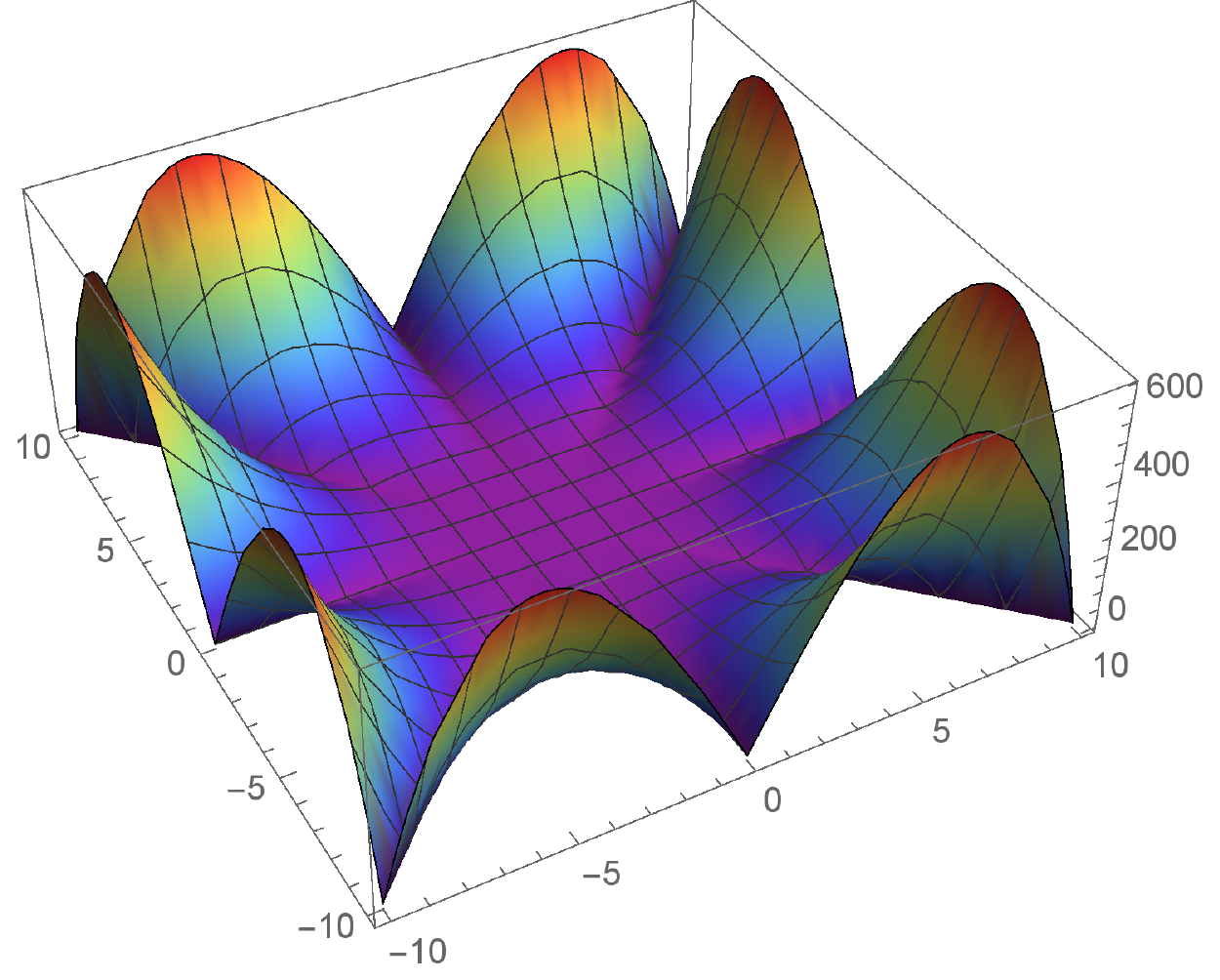}
\end{center}
\caption{[Color online] \textit{The dimension 
$|D(p,q)| = |\frac{1}{6} x y (x - y)(x + y)|$ of an irreducible $so(5) = sp(2)$
representation as a 3-dimensional plot over the $(x,y)$-plane.}}
\label{sp2landscape3d}
\end{figure}

The Cartesian coordinates of a representation $(p,q)$ in the landscape are 
given by
\begin{equation}
x = p + q + 2, \quad y = q + 1.
\end{equation}
The dimension of the representation can then be expressed as
\begin{equation}
D(p,q) = \frac{1}{6} x y (x - y)(x + y) = \frac{1}{6} x y (x^2 - y^2).
\end{equation}
This expression vanishes along the straight lines in Fig.\ref{sp2landscape}
that separate different sectors of the landscape with a 45 degrees opening 
angle. The sign of $D(p,q)$ determines the sign $\pm$ with which representations
in a given sector contribute to tensor product reductions. The landscape is also
illustrated in Fig.\ref{sp2landscape3d} as a 3-dimensional plot of $|D(p,q)|$ 
over the $(x,y)$-plane.

\subsection{Antoine-Speiser scheme for $so(5) = sp(2)$}

We are now prepared to discuss the Antoine-Speiser scheme for $so(5) = sp(2)$.
Just as we illustrated for the simple $su(2)$ case in the Introduction, in order
to perform a tensor product reduction, one superimposes the weight diagram of
the first representation on the landscape, centered at the position of the 
second representation in a positive sector. The states in the weight diagram 
then mark those representations in the landscape that contribute to the 
reduction. Each representation appears with a multiplicity given by the 
degeneracy of the corresponding state in the weight diagram. Recall that for 
$su(2)$ there were no degeneracies and thus each representation had multiplicity
1. In addition, just as for $su(2)$, each representation occurs with a positive 
or negative sign, depending on the sector it is positioned in. States that fall 
on top of the lines separating different sectors do not contribute to the 
reduction. In Fig.\ref{Speisersp2} the Antoine-Speiser scheme is illustrated 
for the tensor product reduction
\begin{eqnarray}
\{40\} \times \{16\}&=&\{154\} + \{105\} + 2 \{81\} + \{55\} + 3 \{35\}
+ 2 \{35'\} + 2 \{30\} + 3 \{14\} \nonumber \\
&+&3 \{10\} + 3 \{5\} + 2 \{1\} \nonumber \\
&-&\{35\} - \{30\} - 2 \{14\} - \{10\} - 2 \{5\}- \{1\} \nonumber \\
&-&\{35'\} - \{10\} - \{1\} \nonumber \\
&=&\{154\} + \{105\} + 2 \{81\} + \{55\} + 2 \{35\}
+ \{35'\} + \{30\} + \{14\} \nonumber \\
&+&\{10\} + \{5\}. 
\end{eqnarray}
Note that the product of the representations $\{40\}$ and  $\{16\}$, which both
have non-trivial duality, results in a sum of representations which all have
trivial duality.

Up to now, we have assumed that the degeneracies of the various states in a
weight diagram are known, but we still need to explain how the degeneracy 
factors are calculated. These factors depend only on $p$ and $q$, i.e.\ on the
shape of the weight diagram of the representation $(p,q)$. In order to 
determine the degeneracy factors we apply the Antoine-Speiser scheme to the
tensor product reduction of the representation $(p,q)$ with the trivial
representation $(0,0) = \{1\}$, which simply results in $(p,q)$ itself. When we
center the weight diagram of the representation $(p,q)$ on the point 
corresponding to $(0,0) = \{1\}$ in the landscape, a number of representations 
in the landscape are covered by states in the weight diagram. Ultimately, only 
the point $(p,q)$ contributes to the tensor product reduction, provided that all
degeneracy factors are properly taken into account. Actually, this requirement 
alone completely determines these factors. In order to calculate the 
degeneracies, one applies a recursive procedure starting from the state in the 
weight diagram that covers the point $(p,q)$ in the landscape. This 
representation occurs only in the positive sector and thus the corresponding 
state in the weight diagram is not degenerate. Using this as well as the 8-fold
symmetry of the weight diagram, one can determine the other degeneracy factors 
by proceeding to other representations in the landscape, which may be covered 
by states in different sectors. The fact that they do not contribute to the 
tensor product reduction uniquely determines the corresponding degeneracy 
factor.

\subsection{Numerical implementation of the Antoine-Speiser \\ method}

Using the weight diagrams of Figs.\ref{sp2weighta} and \ref{sp2weightb} as well
as the landscape of Fig.\ref{sp2landscape}, one can reduce a large number of 
tensor products by hand. In order to automate this process and in order to
access even larger representations, we have developed a corresponding FORTRAN 
code. After specifying the two tensor product partners by $(p_1,q_1)$ and 
$(p_2,q_2)$, the degeneracies in the weight diagram of the representation 
$(p_1,q_1)$ are determined recursively, by demanding that 
$(p_1,q_1) \times (0,0) = (p_1,q_1)$. The resulting weight diagram is then 
superimposed on the landscape, centered at the position $(p_2,q_2)$, and the 
contributions to the tensor product are identified. In this way, one can
calculate, for example, the tensor product reduction of $(9,9) = \{10000\}$ and 
$(2,0) = \{10\}$, which results in
\begin{eqnarray}
\{10000\} \times \{10\}&=&\{14080\} + \{12320\} + \{11340\} + \{10240\} +
2 \{10000\} \nonumber \\
&+&\{8960\} + \{8360\} + \{7980\} + \{6720\}. 
\end{eqnarray}
The corresponding output of the FORTRAN code then looks as follows
\begin{eqnarray}
&&( 9, 9)*( 2, 0) =  \{10000\}* \{10\} \nonumber \\
&&1( 11, 9) =  1 \{14080\} \nonumber \\
&&1( 9, 10) =  1 \{12320\} \nonumber \\
&&1( 11, 8) =  1 \{11340\} \nonumber \\
&&1( 7, 11) =  1 \{10240\} \nonumber \\
&&2( 9, 9) =  2 \{10000\} \nonumber \\
&&1( 11, 7) =  1 \{8960\} \nonumber \\
&&1( 7, 10) =  1 \{8360\} \nonumber \\
&&1( 9, 8) =  1 \{7980\} \nonumber \\
&&1( 7, 9) =  1 \{6720\}
\end{eqnarray}
The code is available from the authors upon request, for the three rank 2 Lie
algebras $so(5) = sp(2)$, $su(3)$, and $g(2)$.

\subsection{Girdle method for $so(5) = sp(2)$ tensor product reduction}

Finally, let us also consider the girdle method. As in the $su(2)$ case, the 
girdle polynomial of an irreducible representation $(p,q)$ with $x = p + q + 2$ 
and $y = q + 1$ is determined by its eight positions $(\pm x,\pm y)$ and 
$(\pm y,\pm x)$ in the positive and negative sectors of the landscape, such that
\begin{eqnarray}
\xi(p,q)&=&x^{p+q+2} y^{q+1} - x^{q+1} y^{p+q+2} + \frac{1}{x^{q+1}} y^{p+q+2}
- \frac{1}{x^{p+q+2}} y^{q+1} \nonumber \\
&+&\frac{1}{x^{p+q+2} y^{q+1}} - \frac{1}{x^{q+1} y^{p+q+2}} 
+ x^{q+1} \frac{1}{y^{p+q+2}} - x^{p+q+2} \frac{1}{y^{q+1}}.
\end{eqnarray}
The girdles of the representations $(0,0) = \{1\}$, $(1,0) = \{4\}$, 
$(0,1) = \{5\}$, and $(1,1) = \{16\}$ are hence given by
\begin{eqnarray}
\xi(0,0)&=&x^2 y - x y^2 + \frac{1}{x} y^2 - \frac{1}{x^2} y 
+ \frac{1}{x^2 y} - \frac{1}{x y^2} + x \frac{1}{y^2} - x^2 \frac{1}{y},
\nonumber \\
\xi(1,0)&=&x^3 y - x y^3 + \frac{1}{x} y^3 - \frac{1}{x^3} y 
+ \frac{1}{x^3 y} - \frac{1}{x y^3} + x \frac{1}{y^3} - x^3 \frac{1}{y},
\nonumber \\
\xi(0,1)&=&x^3 y^2 - x^2 y^3 + \frac{1}{x^2} y^3 - \frac{1}{x^3} y^2 
+ \frac{1}{x^3 y^2} - \frac{1}{x^2 y^3} + x^2 \frac{1}{y^3} - x^3 \frac{1}{y^2},
\nonumber \\
\xi(1,1)&=&x^4 y^2 - x^2 y^4 + \frac{1}{x^2} y^4 - \frac{1}{x^4} y^2 
+ \frac{1}{x^4 y^2} - \frac{1}{x^2 y^4} + x^2 \frac{1}{y^4} - x^4 \frac{1}{y^2}.
\end{eqnarray}
The characteristic Laurent polynomial of the irreducible representation $(p,q)$
is again given by
\begin{equation}
\chi(p,q) = \frac{\xi(p,q)}{\xi(0,0)}.
\end{equation}
In the graphical method, this corresponds to determining the degeneracies in
the weight diagram of the irreducible representation $(p,q)$ by superimposing 
it at the position of $(0,0)$ in the landscape. For example, for the 
fundamental spinor representation $(1,0) = \{4\}$ and for the vector 
representation $(0,1) = \{5\}$ one obtains
\begin{eqnarray}
\chi(1,0)&=&\frac{\xi(1,0)}{\xi(0,0)} = x + y + \frac{1}{x} + \frac{1}{y},
\nonumber \\
\chi(0,1)&=&\frac{\xi(0,1)}{\xi(0,0)} = 
x y + \frac{1}{x} y + \frac{1}{x y} + x \frac{1}{y} + 1,
\end{eqnarray}
which are indeed the characteristic Laurent polynomials associated with the
corresponding weight diagrams in Figs.\ref{sp2weighta} and \ref{sp2weightb}.

The decomposition of the tensor product of the two representations $(p_1,q_1)$ 
and $(p_2,q_2)$ into irreducible representations $(p,q)$,
\begin{equation}
(p_1,q_1) \times (p_2,q_2) = \sum_{p,q} n(p_1,q_1;p_2,q_2;p,q) (p,q),
\end{equation} 
then results from
\begin{equation}
\chi(p_1,q_1) \xi(p_2,q_2) = \sum_{p,q} n(p_1,q_1;p_2,q_2;p,q) \xi(p,q),
\end{equation}
where $n(p_1,q_1;p_2,q_2;p,q)$ denotes the multiplicity of the representation
$(p,q)$. Unlike for $su(2)$, $n(p_1,q_1;p_2,q_2;p,q)$ is now no longer limited
to 0 or 1. In the graphical method, this corresponds to superimposing the 
weight diagram of the irreducible representation $(p_1,q_1)$ at the position of 
$(p_2,q_2)$ in the landscape. The decomposition of the tensor product of the 
two representations $(0,1)$ and $(1,0)$ then results from
\begin{eqnarray}
\chi(0,1) \xi(1,0)&=&
\left(x y + \frac{1}{x} y + \frac{1}{x y} + x \frac{1}{y} + 1\right) 
\nonumber \\
&\times&\left(x^3 y - x y^3 + \frac{1}{x} y^3 - \frac{1}{x^3} y 
+ \frac{1}{x^3 y} - \frac{1}{x y^3} + x \frac{1}{y^3} - x^3 \frac{1}{y}\right)
\nonumber \\
&=&x^4 y^2 - x^2 y^4 + y^4 - \frac{1}{x^2} y^2 
+ \frac{1}{x^2} - \frac{1}{y^2} + x^2 \frac{1}{y^2} - x^4
\nonumber \\
&+&x^2 y^2 - y^4 + \frac{1}{x^2} y^4 - \frac{1}{x^4} y^2 
+ \frac{1}{x^4} - \frac{1}{x^2 y^2} + \frac{1}{y^2} - x^2 
\nonumber \\
&+&x^2 - y^2 + \frac{1}{x^2} y^2 - \frac{1}{x^4} 
+ \frac{1}{x^4 y^2} - \frac{1}{x^2 y^4} + \frac{1}{y^4} - x^2 \frac{1}{y^2}
\nonumber \\
&+&x^4 - x^2 y^2 + y^2 - \frac{1}{x^2} 
+ \frac{1}{x^2 y^2} - \frac{1}{y^4} + x^2 \frac{1}{y^4} - x^4 \frac{1}{y^2}
\nonumber \\
&+&x^3 y - x y^3 + \frac{1}{x} y^3 - \frac{1}{x^3} y 
+ \frac{1}{x^3 y} - \frac{1}{x y^3} + x \frac{1}{y^3} - x^3 \frac{1}{y}
\nonumber \\
&=&\xi(1,0) + \xi(1,1).
\end{eqnarray}
Hence, we have obtained
\begin{eqnarray}
&&(0,1) \times (1,0) = (1,0) + (1,1) \Rightarrow \nonumber \\
&&\{5\} \times \{4\} = \{4\} + \{16\}.
\end{eqnarray}
Even for this rather simple problem, the algebraic girdle method is much more 
tedious than the graphical Antoine-Speiser scheme.

\section{Tensor Product Reduction for $su(3)$}

In this section, we develop the Antoine-Speiser scheme for the algebra $su(3)$.
Again, after defining the algebra and the corresponding Lie group, we consider 
the weight diagrams of several irreducible representations as well as the 
corresponding landscape. The method for tensor product reduction then works
exactly as in the $so(5) = sp(2)$ case.

\subsection{The unitary group $SU(n)$ and its algebra $su(n)$}

The unitary $n \times n$ matrices with determinant 1 form a group under matrix 
multiplication --- the special unitary group $SU(n)$. Each element 
$U \in SU(n)$ can be represented as
\begin{equation}
U = \exp(i H),
\end{equation}
where $H$ is Hermitean and traceless. The matrices $H$ form the $su(n)$ 
algebra. It has $n^2-1$ free parameters, and hence $n^2-1$ generators $T^a$,
among which $n-1$ commute with each other. Thus the rank of $su(n)$ is 
$n - 1$. The simplest non-trivial representations of $su(n)$ are the fundamental
$n$-dimensional representation $\{n\}$ and its conjugate representation 
$\{\overline n\}$. For $su(2)$ the conjugate representation $\{\overline 2\}$ 
is unitarily equivalent to $\{2\}$ which thus is pseudo-real. This is not the
case for $su(n)$ with $n \geq 3$ which also has complex representations.

The center of $SU(n)$ is the group 
$\Z(n) = \{\exp(2 \pi i m/n) \1, m = 0,1,\dots,n-1\}$ consisting of the 
unit-matrix $\1$ multiplied by a complex $n$-th root $\exp(2 \pi i m/n)$ of 1. 
These matrices obviously commute with all other group elements. In addition, 
they are unitary and have determinant 1, and thus indeed belong to $SU(n)$.
The group manifold of $SU(n)$ is locally a product of spheres
\begin{equation}
\label{SUnmanifold}
SU(n) = S^3 \times S^5 \times \dots \times S^{2n-1}.
\end{equation}

\subsection{Weight diagrams of $su(3)$ representations}

\begin{figure}[tbh]
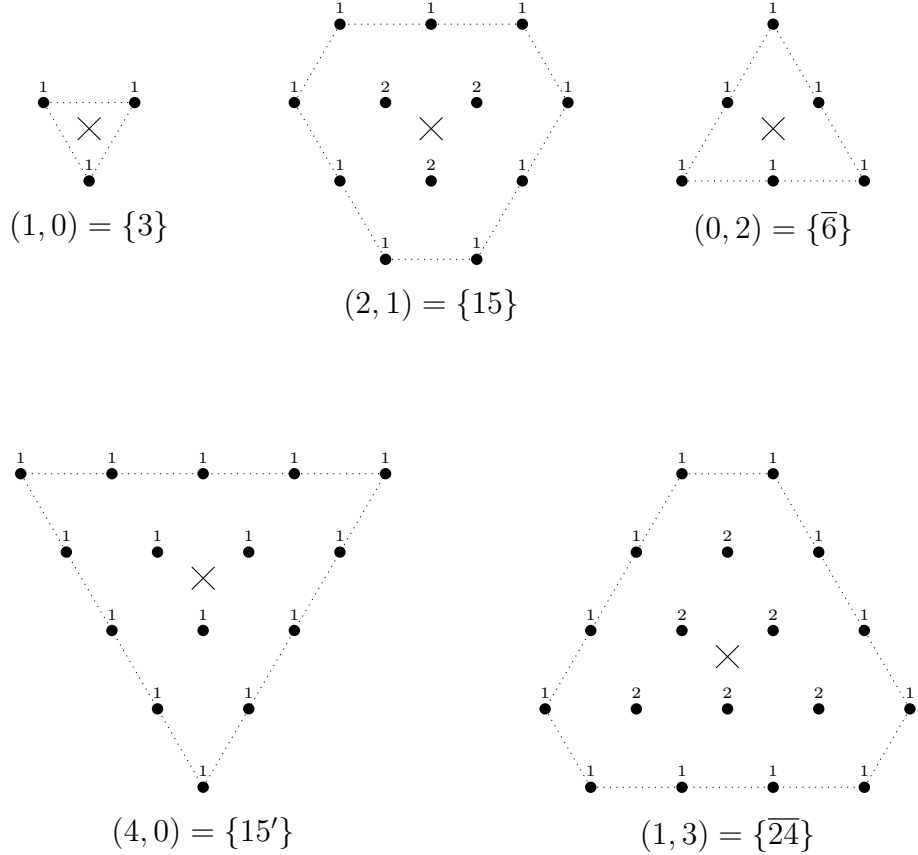

\begin{center}

%

\end{center}
\caption{\it The weight diagrams of several $su(3)$ representations of the same
non-trivial triality. The axes are labeled by the eigenvalues of the commuting 
diagonal generators $T_3$ and $T_8$ of the 8-dimensional algebra $su(3)$. Note 
that the states in these weight diagrams belong to one of three triangular
sublattices. The superscripts denote the degeneracies of the various states. A 
representation $(p,q)$ (which we alternatively denote as $\{D(p,q)\}$) is 
uniquely characterized by the side lengths $p$ and $q$ of its hexagon-shaped 
weight diagram which determine its dimension $D(p,q)$.}
\label{su3weighta}
\end{figure}

\begin{figure}[tbh]
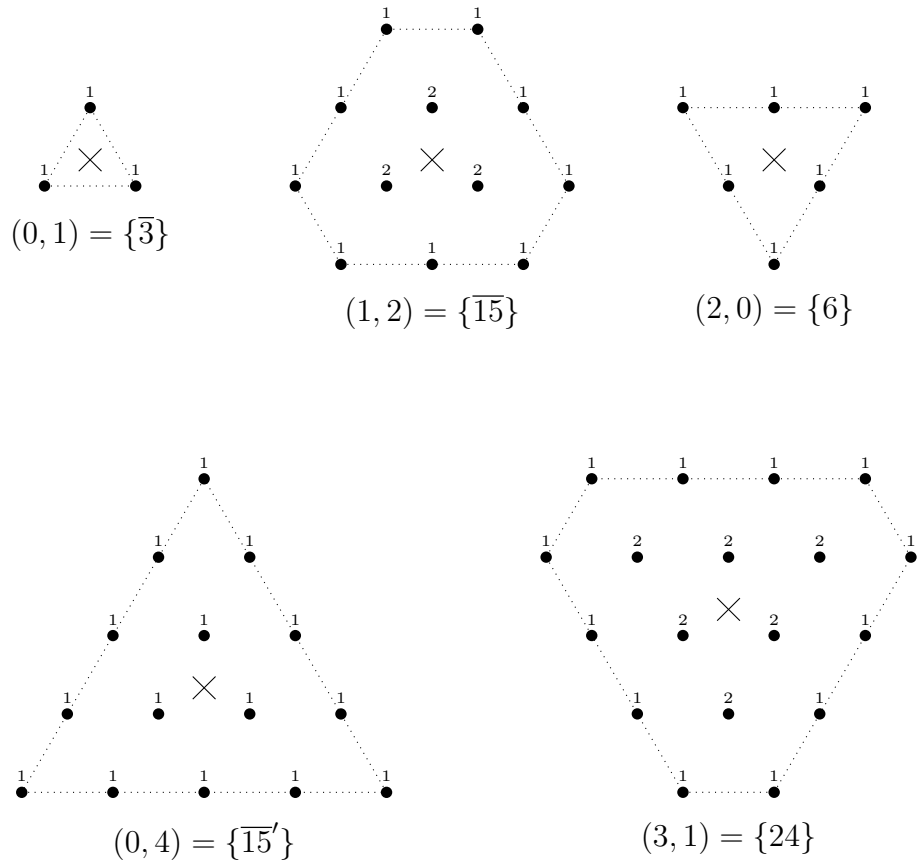

\begin{center}



\end{center}
\caption{\it The weight diagrams of the $su(3)$ representations that are 
conjugate to those in Fig.\ref{su3weighta}. These belong to another triangular
sublattice and thus have the opposite triality.}
\label{su3weightb}
\end{figure}

\begin{figure}[tbh]
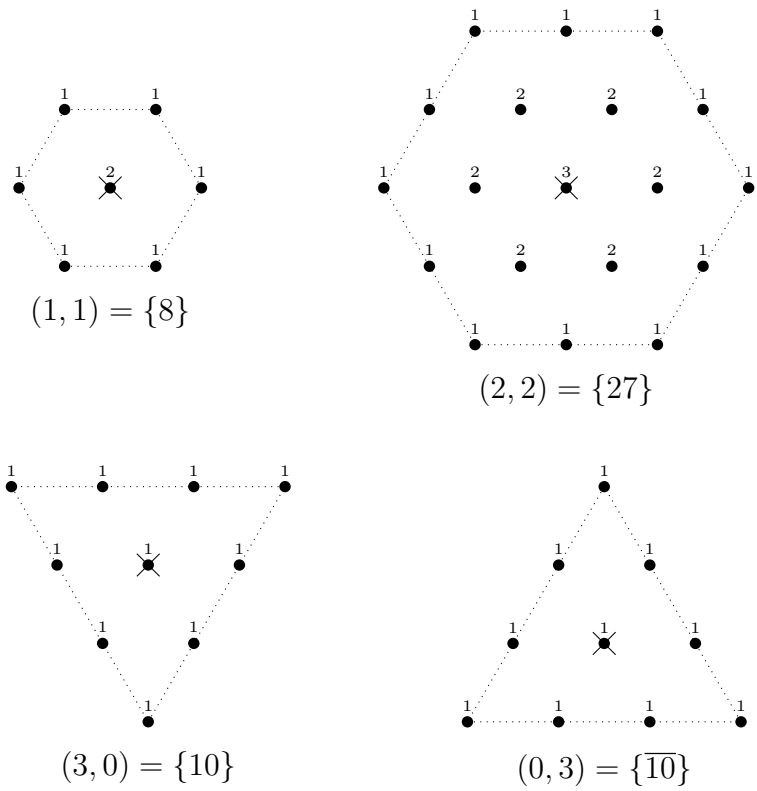

\begin{center}

%

\end{center}
\caption{\it The weight diagrams of several $su(3)$ representations of
trivial triality, which belong to the third triangular sublattice. In this 
case, the origin is always occupied by a state in the weight diagram.}
\label{su3weightc}
\end{figure}

\begin{figure}[tbh]
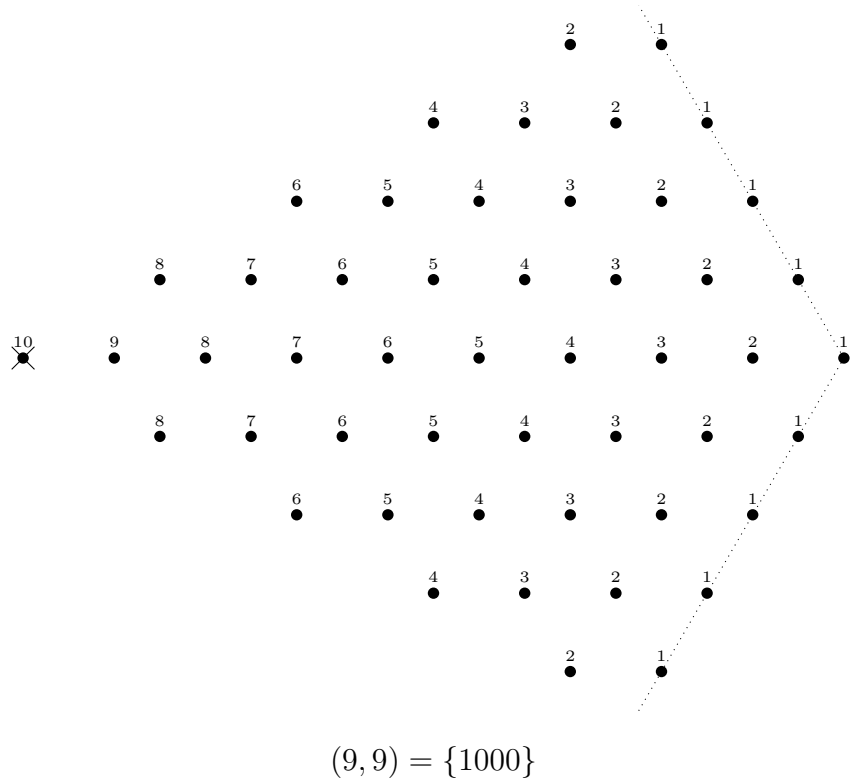

\begin{center}

%

\end{center}
\caption{\it One of six segments of the weight diagram of the $su(3)$ 
representation $(9,9) = \{1000\}$.}
\label{su3WD1000}
\end{figure}

Since $su(3)$ has rank 2, the weight diagrams of its irreducible 
representations can again be drawn in a 2-dimensional plane. The eigenvalues of 
the diagonal generators $T_3$ and $T_8$ characterize the states of an 
irreducible representation. Unlike for $su(2)$ and just as for $so(5) = sp(2)$,
states may again be degenerate, i.e.\ different states may have the same 
eigenvalues of $T_3$ and $T_8$. Since $SU(3)$ has the center $\Z(3)$, there are 
three classes of $su(3)$ representations with different triality. The weight 
diagrams of several complex representations with the same non-trivial triality, 
including the fundamental representation $\{3\}$, are illustrated in 
Fig.\ref{su3weighta}. The points in the weight diagrams of these 
representations belong to one triangular sublattice. The weight diagrams of the 
representations conjugate to those of Fig.\ref{su3weighta}, including the 
anti-fundamental representation $\{\overline{3}\}$,  are shown in 
Fig.\ref{su3weightb}. Their points belong to another triangular sublattice and 
have opposite non-trivial triality. Several representations of trivial 
triality, including the real adjoint representation $\{8\}$ are depicted in 
Fig.\ref{su3weightc}. These representations belong to the third triangular 
sublattice. In this case, the origin is always occupied by a state in the 
weight diagrams. The weight diagram of a general representation has the shape 
of a hexagon characterized by its side lengths $p$ and $q$. The dimension of a 
representation is determined by $p$ and $q$ and is given by
\begin{equation}
\label{Dsu3}
D(p,q) = \frac{1}{2}(p + 1)(q + 1)(p + q + 2).
\end{equation}
While an irreducible representation $(p,q)$ is completely characterized by the 
side lengths $p$ and $q$ of its hexagon-shaped weight diagram, it is again not
uniquely determined by its dimension $D(p,q)$ alone. For example, 
$D(0,4) = D(1,2) = 15$ and $D(0,14) = D(1,9) = D(3,5) = 120$. As before, we 
alternatively denote a representation $(p,q)$ by $\{D(p,q)\}$. To distinguish 
the ambiguous cases, we write $(1,2) = \{15\}$, $(0,4) = \{15'\}$, 
$(3,5) = \{120\}$, $(1,9) = \{120'\}$, and $(0,14) = \{120''\}$. The 
degeneracies for $D(p,q) \leq 1000$ are listed in Table \ref{tablesu3}. For
$D(p,q) \leq 10^8$ one encounters degeneracies as large as $g = 22$ (including
a factor of 2 for the trivial degeneracy of complex conjugate representations),
for example,
\begin{equation}
D(0,383) = D(5,153) = D(7,131) = D(13,95) = D(27,59) = D(39,43) = 73920.
\end{equation}
Again as a side remark, we like to mention that in this case $(9,9) = \{1000\}$.
The weight diagram of this representation is illustrated in Fig.\ref{su3WD1000}.

\begin{table}[tbh]
\begin{center}
\begin{tabular}{|c|c|c|c|c|}
\hline 
$D(p,q)$ & $g$ & $(p_1,q_1)$ & $(p_2,q_2)$ & $(p_3,q_3)$ \\
\hline
\hline
15       &  4  & $(0,4)$     & $(1,2)$    &             \\
\hline
105      &  4  & $(0,13)$    & $(2,6)$    &             \\
\hline
120      &  6  & $(0,14)$    & $(1,9)$    & $(3,5)$     \\
\hline
195      &  4  & $(1,12)$    & $(2,9)$    &             \\
\hline
210      &  4  & $(0,19)$    & $(4,6)$    &             \\
\hline
231      &  4  & $(0,20)$    & $(2,10)$   &             \\
\hline
405      &  4  & $(2,14)$    & $(5,8)$    &             \\
\hline
440      &  4  & $(1,19)$    & $(4,10)$   &             \\
\hline
504      &  4  & $(3,13)$    & $(6,8)$    &             \\
\hline
510      &  4  & $(2,16)$    & $(4,11)$   &             \\
\hline
528      &  4  & $(0,31)$    & $(1,21)$   &             \\
\hline
561      &  4  & $(0,32)$    & $(5,10)$   &             \\
\hline
595      &  4  & $(0,33)$    & $(6,9)$    &             \\
\hline
741      &  4  & $(0,37)$    & $(5,12)$   &             \\
\hline
840      &  6  & $(1,27)$    & $(4,15)$   & $(5,13)$    \\
\hline
960      &  6  & $(1,29)$    & $(3,19)$   & $(7,11)$    \\
\hline
990      &  4  & $(0,43)$    & $(8,10)$   &             \\
\hline
\end{tabular}
\end{center}
\caption{\it Non-trivial degeneracies $g \geq 4$ of the dimensions 
$D(p,q) \leq 1000$ of $su(3)$ representations $(p,q)$. Only complex 
representations (with $p \neq q$) are found to be degenerate. The degeneracy 
factor $g$ includes a factor of 2 due to a trivial degeneracy with the complex 
conjugate representations $\{\overline{D(p,q)}\} = (q,p)$, whose $(q,p)$ values 
are not listed explicitly.}
\label{tablesu3}
\end{table}

As one sees in the weight diagrams of Figs.\ref{su3weighta}, \ref{su3weightb}, 
and \ref{su3weightc}, the degeneracies of the individual states follow a shell
structure. In particular, the states in the outer shell at the edge of the
weight diagram are not degenerate. The states in the next inner shell are
two-fold degenerate. As one moves on to further interior shells, step by step 
the degeneracy increases by one, until the shell reaches a triangular shape. 
From that point on, the degeneracy remains constant and does not increase 
further for the additional interior triangular shells. This behavior reflects 
itself in eq.(\ref{Dsu3}) for $D(p,q)$ which obeys
\begin{eqnarray}
D(p,q)&=&\frac{1}{2} (p + 1)(q + 1)(p + q + 2) \nonumber \\
&=&\frac{1}{2} p q (p + q) + \frac{1}{2}(p^2 + 4 p q + q^2 + 3 p + 3 q + 2)
\nonumber \\
&=&D(p - 1,q - 1) + A(p,q), \qquad p, q \geq 1.
\end{eqnarray}
Here 
\begin{equation}
A(p,q) = \frac{1}{2}(p^2 + 4 p q + q^2 + 3 p + 3 q + 2)
\end{equation}
is again the area of a weight diagram. By subtracting $A(p,q)$ from $D(p,q)$,
as long as $p, q \geq 1$, one removes the outer shell of the weight diagram of 
the representation $(p,q)$ and one reduces the degeneracies of all other 
states by 1, thus arriving at the representation $(p - 1,q - 1)$. Once the 
weight diagram reaches a triangular shape, which is the case for $p = 0$ or 
$q = 0$, one obtains
\begin{equation}
D(p,0) = \frac{1}{2} (p + 1)(p + 2) = A(p,0), \qquad
D(0,q) = \frac{1}{2} (q + 1)(q + 2) = A(0,q).
\end{equation}

\subsection{Landscape of $su(3)$ representations}

Since $su(3)$ also has rank 2, the landscape of its irreducible representations 
can again be drawn in a 2-dimensional plane. As shown in Fig.\ref{su3landscape},
it corresponds to a triangular lattice, which consists of three triangular
sublattices associated with the three trialities.

\begin{figure}[tbp]
\begin{center}
\includegraphics[width=\textwidth]{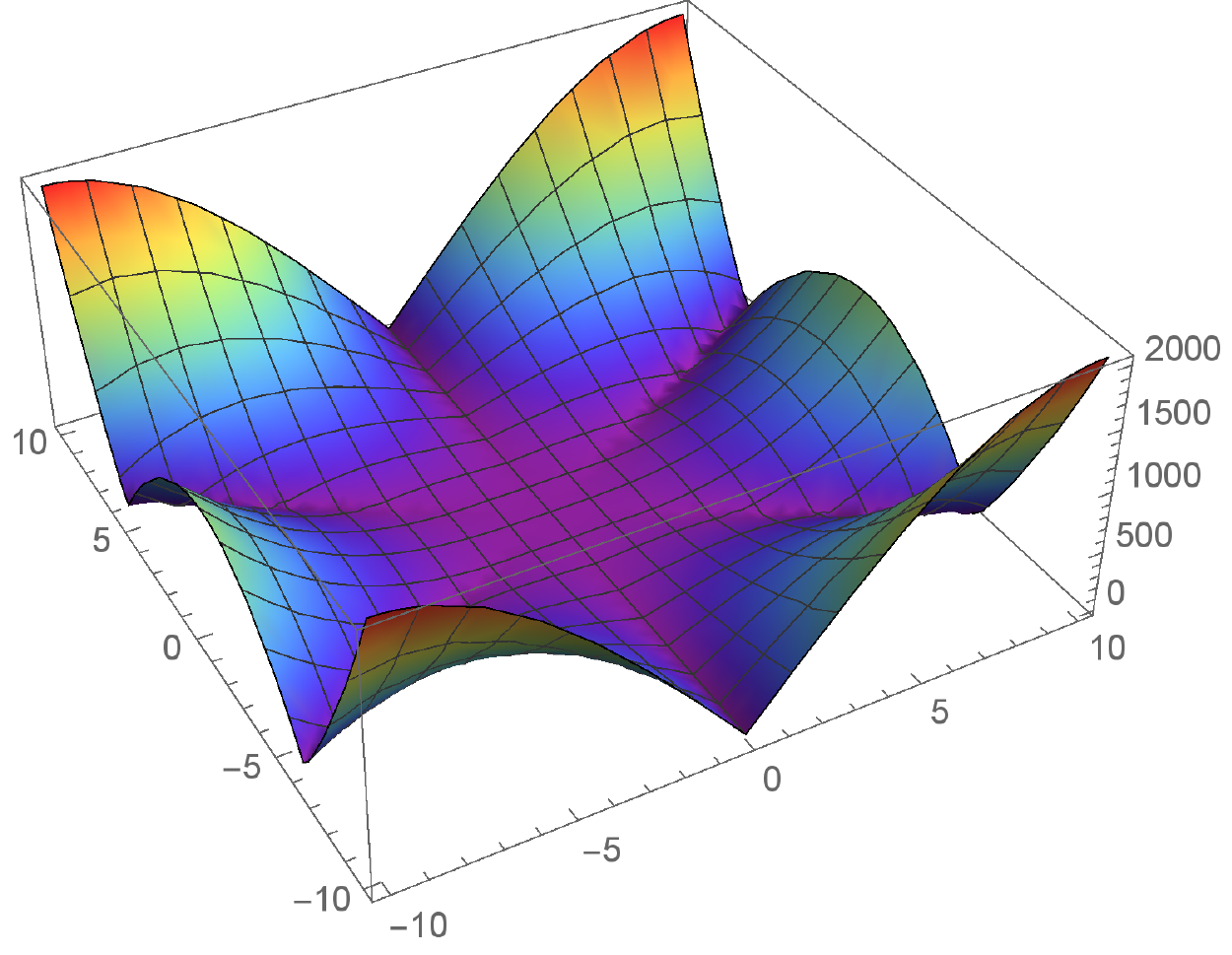}
\end{center}
\caption{[Color online] \textit{The dimension 
$|D(p,q)| = |x (x - \sqrt{3} y)(x + \sqrt{3} y)|$ of an irreducible $su(3)$
representation as a 3-dimensional plot over the $(x,y)$-plane.}}
\label{su3landscape3d}
\end{figure}

The Cartesian coordinates of a representation $(p,q)$ in the landscape are 
given by
\begin{equation}
x = \frac{1}{2}(p + q + 2), \quad y = \frac{1}{2 \sqrt{3}}(p - q).
\end{equation}
The dimension can then be expressed as
\begin{equation}
D(p,q) = x (x - \sqrt{3} y)(x + \sqrt{3} y) = x (x^2 - 3 y^2).
\end{equation}
This expression vanishes along the straight lines in Fig.\ref{su3landscape}
that separate different sectors of the landscape with a 60 degrees opening 
angle. Again, the sign of $D(p,q)$ determines the sign $\pm$ with which 
representations in a given sector contribute to tensor product reductions. We 
again show a 3-dimensional plot of $|D(p,q)|$ over the $(x,y)$-plane in 
Fig.\ref{su3landscape3d}. 

The tensor product reduction works exactly as for $so(5) = sp(2)$ and will thus 
not be discussed again. Also its numerical implementation works as before. In
this way, one obtains, for example,
\begin{eqnarray}
\{1000\} \times \{10\}&=&\{1495\} + \{1331\} + \{1134\} + 
\{\overline{1134}\} + \{1000\} \nonumber \\
&+&\{\overline{910}\} + \{836\} + \{\overline{836}\} + \{729\} + \{595\}. 
\end{eqnarray}

\subsection{Girdle method for $su(3)$ tensor product reduction}

Let us now consider the girdle method for $su(3)$. In this case, the girdle
polynomial of an irreducible representation $(p,q)$ with
\begin{equation} 
\widetilde x = 2 x = p + q + 2, \quad \widetilde y = 2 \sqrt{3} y = p - q
\end{equation}
is determined by its six positions in the positive and negative sectors of the 
landscape, such that
\begin{eqnarray}
\xi(p,q)&=&\widetilde x^{p+q+2} \widetilde y^{p-q} 
- \widetilde x^{p+1} \widetilde y^{p+2q+3} 
+ \frac{1}{\widetilde x^{p+1}} \widetilde y^{p+2q+3} \nonumber \\
&-&\frac{1}{\widetilde x^{p+q+2}} \widetilde y^{p-q}
+ \frac{1}{\widetilde x^{q+1} \widetilde y^{2p+q+3}} 
- \widetilde x^{q+1} \frac{1}{\widetilde y^{2p+q+3}}.
\end{eqnarray}
The girdle of the trivial representation $(0,0) = \{1\}$ is thus given by
\begin{equation}
\xi(0,0) = \widetilde x^2 - \widetilde x \widetilde y^3 
+ \frac{1}{\widetilde x} \widetilde y^3
- \frac{1}{\widetilde x^2} + \frac{1}{\widetilde x \widetilde y^3} 
- \widetilde x \frac{1}{\widetilde y^3},
\end{equation}
while the girdles of the fundamental representations $(1,0) = \{3\}$ and
$(0,1) = \{\overline 3\}$ as well as of the adjoint representation 
$(1,1) = \{8\}$ take the form
\begin{eqnarray}
\xi(1,0)&=&\widetilde x^3 \widetilde y
- \widetilde x^2 \widetilde y^4 + \frac{1}{\widetilde x^2} \widetilde y^4 
- \frac{1}{\widetilde x^3} \widetilde y 
+ \frac{1}{\widetilde x \widetilde y^5} 
- \widetilde x \frac{1}{\widetilde y^5}, \nonumber \\
\xi(0,1)&=&\widetilde x^3 \frac{1}{\widetilde y} - \widetilde x \widetilde y^5 
+ \frac{1}{\widetilde x} \widetilde y^5 
- \frac{1}{\widetilde x^3 \widetilde y} 
+ \frac{1}{\widetilde x^2 \widetilde y^4} 
- \widetilde x^2 \frac{1}{\widetilde y^4}, \nonumber \\
\xi(1,1)&=&\widetilde x^4 - \widetilde x^2 \widetilde y^6
+ \frac{1}{\widetilde x^2} \widetilde y^6 - \frac{1}{\widetilde x^4}
+ \frac{1}{\widetilde x^2 \widetilde y^6} 
- \widetilde x^2 \frac{1}{\widetilde y^6}.
\end{eqnarray}
As before, the characteristic Laurent polynomial of the irreducible 
representation $(p,q)$ is given by
\begin{equation}
\chi(p,q) = \frac{\xi(p,q)}{\xi(0,0)}.
\end{equation}
Hence, for the fundamental representations $(1,0) = \{3\}$ and 
$(0,1) = \{\overline 3\}$ one obtains
\begin{eqnarray}
&&\chi(1,0) = \frac{\xi(1,0)}{\xi(0,0)} = \widetilde x \widetilde y +
\frac{1}{\widetilde x} \widetilde y + \frac{1}{\widetilde y^2}, \nonumber \\
&&\chi(0,1) = \frac{\xi(0,1)}{\xi(0,0)} = \widetilde x \frac{1}{\widetilde y} +
\frac{1}{\widetilde x \widetilde y} + \widetilde y^2.
\end{eqnarray}
These are indeed the characteristic Laurent polynomials associated with the
corresponding weight diagrams in Figs.\ref{su3weighta} and \ref{su3weightb}. 
The decomposition of the tensor product of the two representations $(1,0)$ and 
$(0,1)$ then results from
\begin{eqnarray}
\chi(1,0) \xi(0,1)&=&\left(\widetilde x \widetilde y +
\frac{1}{\widetilde x} \widetilde y + \frac{1}{\widetilde y^2}\right)
\left(\widetilde x^3 \frac{1}{\widetilde y} - \widetilde x \widetilde y^5 
+ \frac{1}{\widetilde x} \widetilde y^5 
- \frac{1}{\widetilde x^3 \widetilde y} 
+ \frac{1}{\widetilde x^2 \widetilde y^4} 
- \widetilde x^2 \frac{1}{\widetilde y^4}\right) \nonumber \\
&=&\widetilde x^4 - \widetilde x^2 \widetilde y^6 + \widetilde y^6 
- \frac{1}{\widetilde x^2} + \frac{1}{\widetilde x \widetilde y^3} 
- \widetilde x^3 \frac{1}{\widetilde y^3} \nonumber \\
&+&\widetilde x^2 - \widetilde y^6 + \frac{1}{\widetilde x^2} \widetilde y^6 
- \frac{1}{\widetilde x^4} + \frac{1}{\widetilde x^3 \widetilde y^3} 
- \widetilde x \frac{1}{\widetilde y^3} \nonumber \\
&+&\widetilde x^3 \frac{1}{\widetilde y^3} - \widetilde x \widetilde y^3 
+ \frac{1}{\widetilde x} \widetilde y^3 
- \frac{1}{\widetilde x^3 \widetilde y^3} 
+ \frac{1}{\widetilde x^2 \widetilde y^6} 
- \widetilde x^2 \frac{1}{\widetilde y^6} \nonumber \\
&=&\xi(0,0) + \xi(1,1).
\end{eqnarray}
Hence, we have obtained
\begin{eqnarray}
&&(1,0) \times (0,1) = (0,0) + (1,1) \Rightarrow \nonumber \\ 
&&\{3\} \times \{\overline 3\} = \{1\} + \{8\}.
\end{eqnarray}
Again, even for this rather simple problem, the algebraic girdle method is less
efficient than the graphical Antoine-Speiser scheme.

\section{Tensor Product Reduction for $g(2)$}

In analogy to $so(5) = sp(2)$ and $su(3)$, we now work out the Antoine-Speiser
scheme for the exceptional Lie algebra $g(2)$, which contains $su(3)$ as a
subalgebra.

\subsection{The exceptional group $G(2)$ and its algebra $g(2)$}

In this subsection we discuss some basic properties of the Lie group $G(2)$ ---
the simplest among the exceptional groups $G(2)$, $F(4)$, $E(6)$, $E(7)$ and
$E(8)$ --- which do not fit into the main sequences $SU(n)$, $Spin(n)$, and 
$Sp(n)$. The group $G(2)$ is interesting because it has a trivial center and 
still is its own universal covering group.

It is natural to construct $G(2)$ as a subgroup of $SO(7)$ which has rank 3 and
21 generators. The $7 \times 7$ real orthogonal matrices $O$ of the group 
$SO(7)$ have determinant 1 and obey the constraint
\begin{equation}
O_{ab} O_{ac} = \delta_{bc}.
\end{equation}
The $G(2)$ subgroup contains those $SO(7)$ matrices that, in addition, satisfy 
the cubic constraint
\begin{equation}
\label{cubic}
T_{abc} = T_{def} O_{da} O_{eb} O_{fc}.
\end{equation}
Here $T$ is a totally anti-symmetric tensor whose non-zero elements follow by 
anti-symmetrization from
\begin{equation}
\label{tensor}
T_{127} = T_{154} = T_{163} = T_{235} = T_{264} = T_{374} = T_{576} = 1.
\end{equation}
Eq.(\ref{tensor}) implies that eq.(\ref{cubic}) represents 7 non-trivial
constraints which reduce the 21 degrees of freedom of $SO(7)$ to the 14 
parameters of $G(2)$. It should be noted that $G(2)$ inherits the reality
properties of $SO(7)$: all its representations are real.

The group manifold of $G(2)$ is the product of the group manifold of $SU(3)$
with a 6-dimensional sphere $S^6$, i.e.\
\begin{equation}
G(2) = SU(3) \times S^6 = S^3 \times S^5 \times S^6.
\end{equation}
From this one obtains
\begin{equation}
SO(7) = S^1 \times S^2 \times S^3 \times S^4 \times S^5 \times S^6 =
G(2) \times S^1 \times S^2 \times S^4 = G(2) \times S^3 \times S^4 = 
G(2) \times S^7.
\end{equation}

\subsection{Weight diagrams of $g(2)$ representations}

\begin{figure}[tbh]
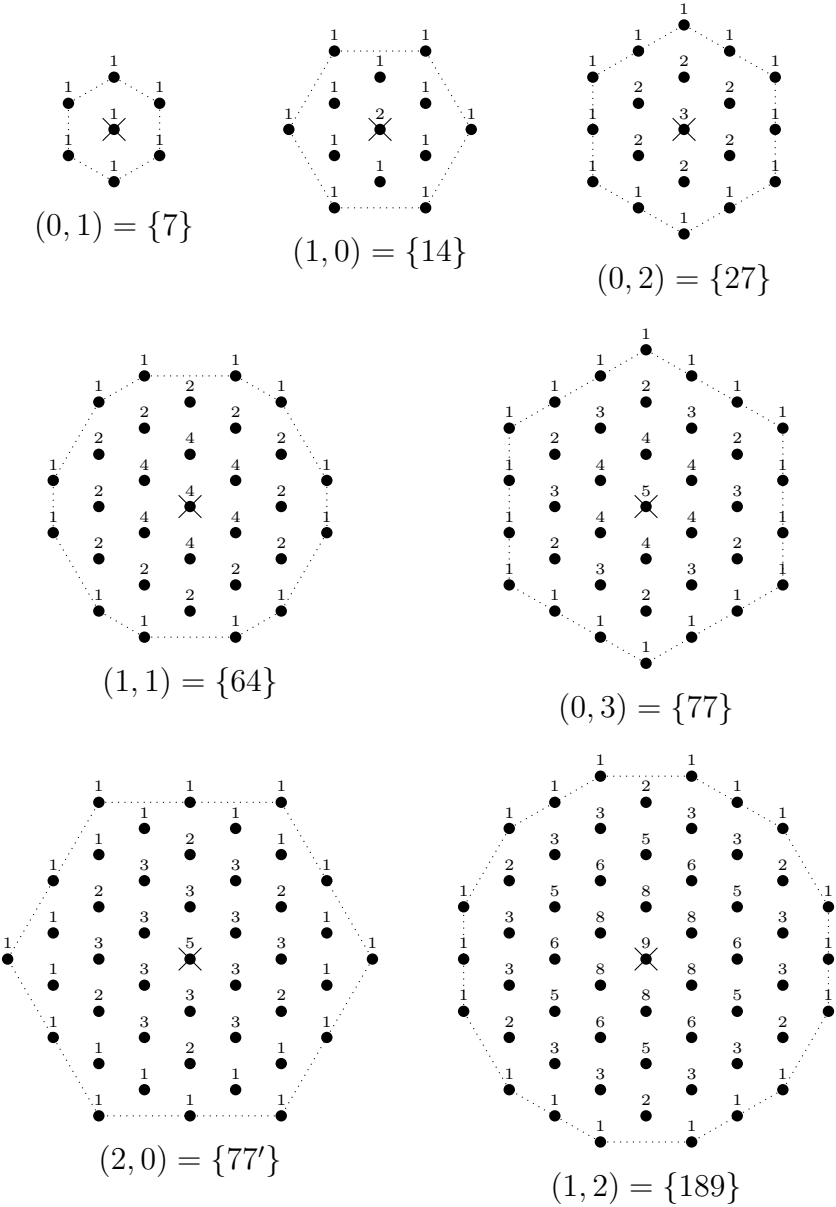

\begin{center}



\end{center}
\caption{\it The weight diagrams of the smallest $g(2)$ representations. The 
axes are labeled by the eigenvalues of the commuting generators $T_3$ and $T_8$ 
of the subalgebra $su(3)$. The superscripts denote the degeneracies of the 
various states. A representation $(p,q)$ (which we alternatively denote as 
$\{D(p,q)\}$) is uniquely characterized by the side lengths $p$ (along the 
$x$-axis) and $q$ (along the $y$-axis) of its dodecagon-shaped weight diagram, 
that determine its dimension $D(p,q)$.}
\label{g2weighta}
\end{figure}

\begin{figure}[tbh]
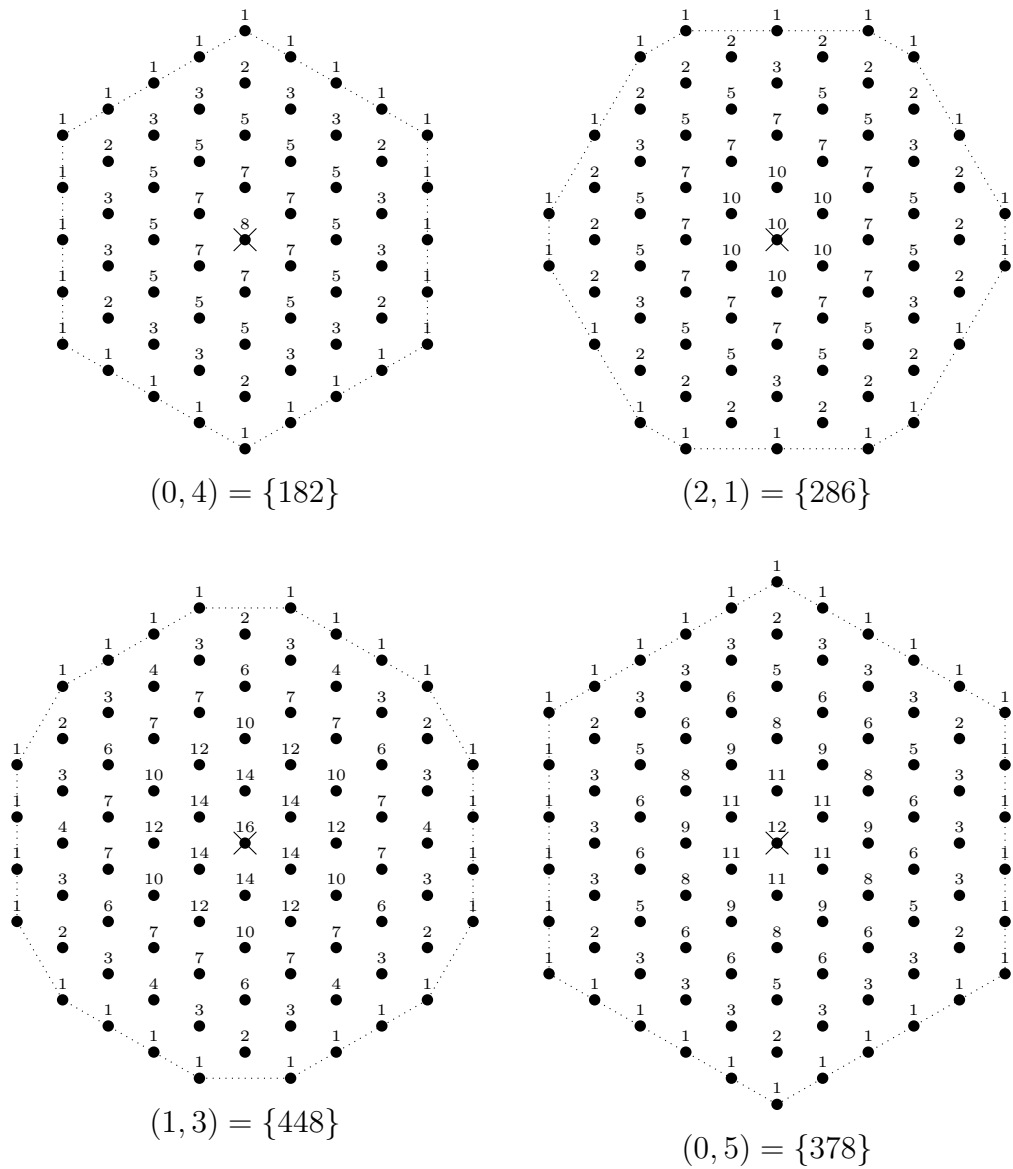

\begin{center}



\end{center}
\caption{\it The weight diagrams of some larger $g(2)$ representations.}
\label{g2weightb}
\end{figure}

\begin{figure}[tbh]
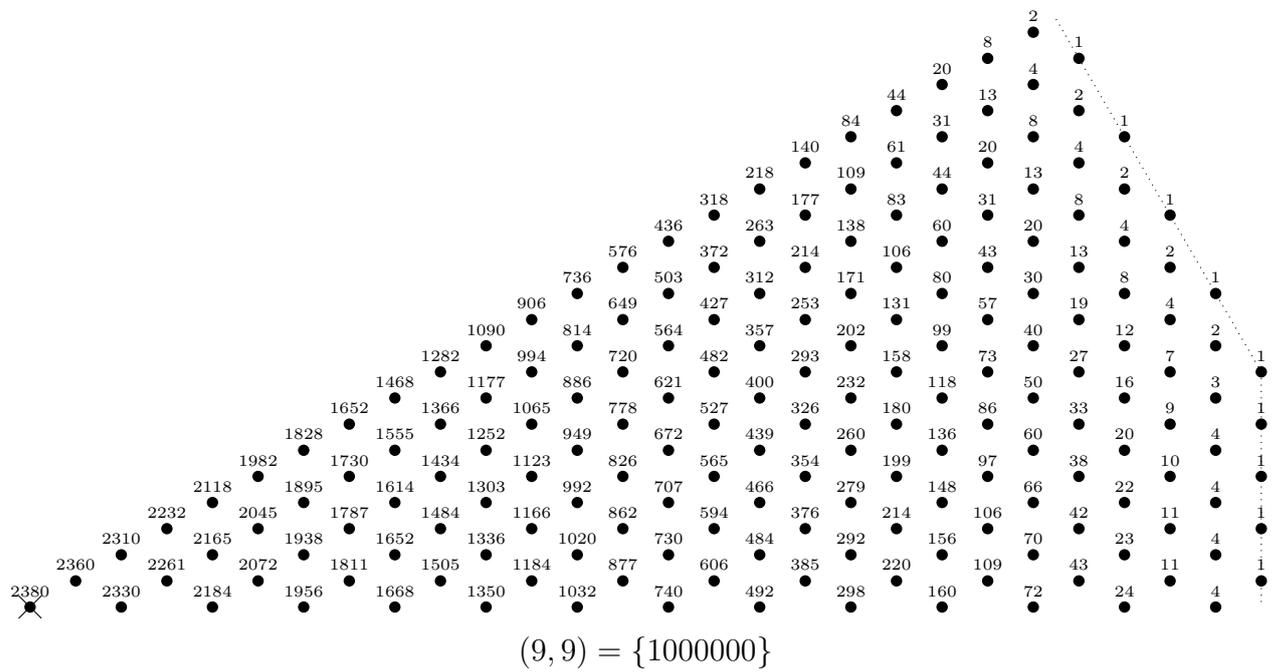

\begin{center}

%

\end{center}
\caption{\it One of twelve segments of the weight diagram of the $g(2)$ 
representation $(9,9) = \{1000000\}$.}
\label{g2WD1000000}
\end{figure}

Since $g(2)$ also has rank 2, the weight diagrams of its irreducible 
representations are again 2-dimensional. Since $su(3)$ is a subalgebra of 
$g(2)$, we can once again use the eigenvalues of the diagonal generators $T_3$ 
and $T_8$ to characterize the states of an irreducible representation. Since 
$G(2)$ has a trivial center, the $SU(3)$ concept of triality does not extend to 
$G(2)$. Several weight diagrams of $g(2)$ representations are shown in 
Figs.\ref{g2weighta} and \ref{g2weightb}. As one sees in Fig.\ref{g2weighta},
under the $su(3)$ subalgebra the fundamental $\{7\}$ representation decomposes
as
\begin{equation}
\label{dec7}
\{7\} = \{3\} + \{\overline 3\} + \{1\},
\end{equation}
while the adjoint $\{14\}$ representation decomposes as
\begin{equation}
\label{dec14}
\{14\} = \{8\} + \{3\} + \{\overline 3\}.
\end{equation}
In both cases, the $g(2)$ representation decomposes into $su(3)$ representations
of three different trialities, which confirms that $G(2)$ indeed has a trivial
center. The weight diagram of a general $g(2)$ representation has the shape of 
a dodecagon characterized by its side lengths $p$ (along the $x$-axis) and $q$. 
The dimension of the representation $(p,q)$ is given by
\begin{equation}
\label{Dg2}
D(p,q) = \frac{1}{120}(p + 1)(q + 1)(p + q + 2)(2p + q + 3)(3p + q + 4)
(3p + 2q + 5).
\end{equation}
In order to distinguish ambiguous cases, we denote $(2,0) = \{77\}$ and 
$(0,3) = \{77'\}$, as well as $(3,2) = \{2079\}$ and $(0,8) = \{2079'\}$.
The degeneracies for $D(p,q) \leq 10^7$ are listed in Table \ref{tableg2}. 
As a side remark, we like to mention that here $(9,9) = \{1000000\}$. The weight
diagram of this representation is illustrated in Fig.\ref{g2WD1000000}.

\begin{table}[tbh]
\begin{center}
\begin{tabular}{|c|c|c|c|}
\hline 
$D(p,q)$ & $g$ & $(p_1,q_1)$ & $(p_2,q_2)$ \\
\hline
\hline
77       &  2  & $(0,3)$     & $(2,0)$    \\
\hline
2079     &  2  & $(0,8)$     & $(3,2)$    \\
\hline
4928     &  2  & $(1,7)$     & $(5,1)$    \\
\hline
30107    &  2  & $(5,4)$     & $(10,0)$   \\
\hline
56133    &  2  & $(2,11)$    & $(8,2)$    \\
\hline
133056   &  2  & $(1,17)$    & $(7,5)$    \\
\hline
315392   &  2  & $(3,15)$    & $(11,3)$   \\
\hline
812889   &  2  & $(0,32)$    & $(4,17)$   \\
\hline
1203125  &  2  & $(4,19)$    & $(14,4)$   \\
\hline
1515591  &  2  & $(2,26)$    & $(11,8)$   \\
\hline
1926848  &  2  & $(11,9)$    & $(21,1)$   \\
\hline
3592512  &  2  & $(5,23)$    & $(17,5)$   \\
\hline
8515584  &  2  & $(3,35)$    & $(15,11)$  \\
\hline
9058973  &  2  & $(6,27)$    & $(20,6)$   \\
\hline
\end{tabular}
\end{center}
\caption{\it Degeneracies $g \geq 2$ of the dimensions $D(p,q) \leq 10^7$ of 
$g(2)$ representations $(p,q)$. The degeneracy factor $g$ counts in how many 
ways $D(p,q)$ can be realized by pairs $(p,q)$.}
\label{tableg2}
\end{table}

\subsection{Landscape of $g(2)$ representations}

\begin{figure}[tbh]
\begin{center}
\includegraphics[width=\textwidth]{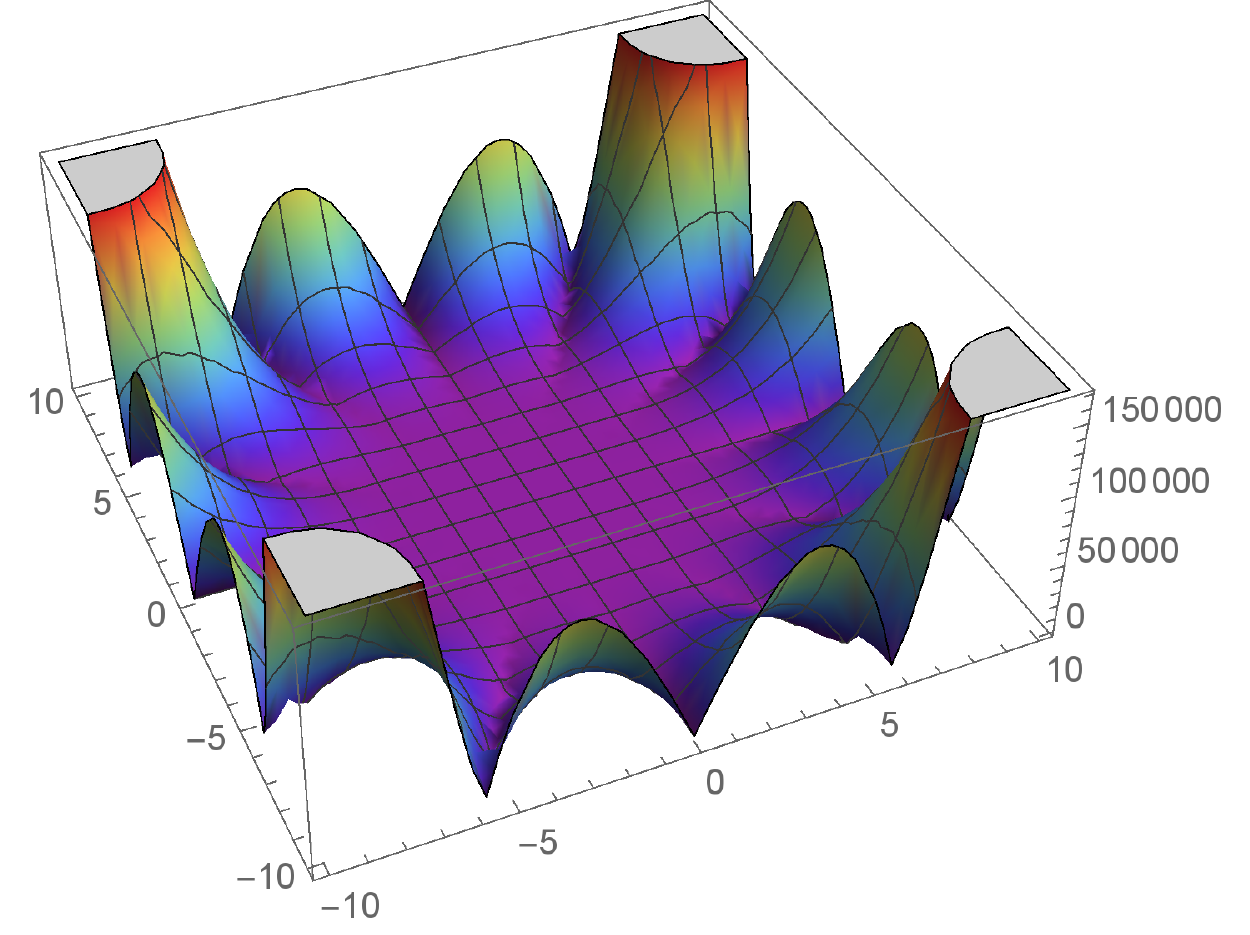}
\end{center}
\caption{[Color online] \textit{The dimension 
$|D(p,q)|= |\frac{1}{10 \sqrt{3}} x y (3 x^4 - 10 x^2 y^2 + 3 y^4)|$ of an 
irreducible $g(2)$ representation as a 3-dimensional plot over the 
$(x,y)$-plane.}}
\label{g2landscape3d}
\end{figure}

As shown in Fig.\ref{g2landscape}, for $g(2)$ the landscape again corresponds 
to a triangular lattice. The Cartesian coordinates of a representation $(p,q)$ 
in the landscape are given by
\begin{equation}
x = \frac{\sqrt{3}}{2}(2 p + q + 3), \quad y = \frac{1}{2}(q + 1).
\end{equation}
The dimension can then be expressed as
\begin{eqnarray}
D(p,q)&=&\frac{1}{10 \sqrt{3}} x y (x - \sqrt{3} y)(x + \sqrt{3} y)
(\sqrt{3} x - y)(\sqrt{3} x + y) \nonumber \\
&=&\frac{1}{10 \sqrt{3}} x y (3 x^4 - 10 x^2 y^2 + 3 y^4).
\end{eqnarray}
This expression again vanishes along the straight lines in Fig.\ref{g2landscape}
that separate twelve different sectors of the landscape, each with a 30 degrees 
opening angle. As before, the sign of $D(p,q)$ determines the sign $\pm$ with 
which representations in a given sector contribute to tensor product reductions.
Finally, we also show a 3-dimensional plot of $|D(p,q)|$ over the $(x,y)$-plane 
in Fig.\ref{g2landscape3d}. 

The tensor product reduction works again as for $so(5) = sp(2)$ and $su(3)$ and 
will not be discussed again. Applying the numerical implementation one obtains, 
for example,
\begin{eqnarray}
\{1000000\} \times \{7\}&=&\{1272271\} + \{1095633\} + \{1127763\} + 
\{1000000\} \nonumber \\
&+&\{839762\} + \{890967\} + \{773604\}. 
\end{eqnarray}

\subsection{Girdle method for $g(2)$ tensor product reduction}

Finally, let us consider the girdle method for $g(2)$. In this case, the girdle
polynomial of an irreducible representation $(p,q)$ with
\begin{equation} 
\widetilde x = \frac{2}{\sqrt{3}} x = 2 p + q + 3, \quad 
\widetilde y = 2 y = q + 1
\end{equation}
is determined by its twelve positions in the positive and negative sectors of
the landscape, such that
\begin{eqnarray}
\xi(p,q)&=&\widetilde x^{2p+q+3} \widetilde y^{q+1} 
- \widetilde x^{p+q+2} \widetilde y^{3p+q+4} 
+ \widetilde x^{p+1} \widetilde y^{3p+2q+5} \nonumber \\
&-&\frac{1}{\widetilde x^{p+1}} \widetilde y^{3p+2q+5}
+ \frac{1}{\widetilde x^{p+q+2}} \widetilde y^{3p+q+4} 
- \frac{1}{\widetilde x^{2p+q+3}} \widetilde y^{q+1} \nonumber \\
&+&\frac{1}{\widetilde x^{2p+q+3} \widetilde y^{q+1}} 
- \frac{1}{\widetilde x^{p+q+2} \widetilde y^{3p+q+4}} 
+ \frac{1}{\widetilde x^{p+1} \widetilde y^{3p+2q+5}} \nonumber \\
&-&\widetilde x^{p+1} \frac{1}{\widetilde y^{3p+2q+5}}
+ \widetilde x^{p+q+2} \frac{1}{\widetilde y^{3p+q+4}} 
- \widetilde x^{2p+q+3} \frac{1}{\widetilde y^{q+1}}.
\end{eqnarray}
The girdle of the trivial representation $(0,0) = \{1\}$ is thus given by
\begin{eqnarray}
\xi(0,0)&=&\widetilde x^3 \widetilde y - \widetilde x^2 \widetilde y^4 
+ \widetilde x \widetilde y^5 - \frac{1}{\widetilde x} \widetilde y^5
+ \frac{1}{\widetilde x^2} \widetilde y^4 
- \frac{1}{\widetilde x^3} \widetilde y \nonumber \\
&+&\frac{1}{\widetilde x^3 \widetilde y} 
- \frac{1}{\widetilde x^2 \widetilde y^4} 
+ \frac{1}{\widetilde x \widetilde y^5}
- \widetilde x \frac{1}{\widetilde y^5}
+ \widetilde x^2 \frac{1}{\widetilde y^4} 
- \widetilde x^3 \frac{1}{\widetilde y}.
\end{eqnarray}
while the girdle of the fundamental representation $(0,1) = \{7\}$ takes the 
form
\begin{eqnarray}
\xi(0,1)&=&\widetilde x^4 \widetilde y^2 - \widetilde x^3 \widetilde y^5 
+ \widetilde x \widetilde y^7 - \frac{1}{\widetilde x} \widetilde y^7
+ \frac{1}{\widetilde x^3} \widetilde y^5 
- \frac{1}{\widetilde x^4} \widetilde y^2 \nonumber \\
&+&\frac{1}{\widetilde x^4 \widetilde y^2} 
- \frac{1}{\widetilde x^3 \widetilde y^5} 
+ \frac{1}{\widetilde x \widetilde y^7}
- \widetilde x \frac{1}{\widetilde y^7}
+ \widetilde x^3 \frac{1}{\widetilde y^5} 
- \widetilde x^4 \frac{1}{\widetilde y^2}.
\end{eqnarray}
As in the other cases, the characteristic Laurent polynomial of the irreducible 
representation $(p,q)$ is given by
\begin{equation}
\chi(p,q) = \frac{\xi(p,q)}{\xi(0,0)}.
\end{equation}
For the fundamental representations $(0,1) = \{7\}$ one then obtains
\begin{equation}
\chi(0,1) = \frac{\xi(0,1)}{\xi(0,0)} = \widetilde x \widetilde y 
+ \widetilde y^2 + \frac{1}{\widetilde x} \widetilde y 
+ \frac{1}{\widetilde x \widetilde y} 
+ \frac{1}{\widetilde y^2} + \widetilde x \frac{1}{\widetilde y} + 1.
\end{equation}
This is indeed the characteristic Laurent polynomial associated with the
corresponding weight diagram in Fig.\ref{g2weighta}.

\section{Conclusions}

We have worked out a graphical tensor product reduction scheme for the simple
rank 2 Lie algebras $so(5) = sp(2)$, $su(3)$, and $g(2)$, which relies on the
fact that 2-dimensional weight diagrams can be superimposed on a 2-dimensional
landscape of irreducible representations. While the method itself extends to
algebras of higher rank, and thus to higher-dimensional weight diagrams and
``landscapes'', in practice the method of superimposing them only works in one
or two dimensions and is thus limited to rank 1 or 2. We have also used the 
algebraic girdle method for tensor product reduction, which is much more 
tedious than the graphical method for calculations by hand even for small 
representations. Furthermore, we have developed computer code for automated 
tensor product reduction, based on the graphical method. This code could be 
extended to higher-rank algebras in a straightforward manner.

\section*{Acknowledgments}

We dedicate this paper to the memory of Petro I.\ Holod, who had a large impact 
as a teacher of group theory and Lie algebras \cite{Hol92},
motivating generations of students to aim at understanding Nature at a deep
level. The research leading to these results has received funding from the 
Schweizerischer Na\-tio\-nal\-fonds and from the European Research Council 
under the European Union's Seventh Framework Programme (FP7/2007-2013)/ ERC 
grant agreement 339220.

\vspace{2cm}

\begin{figure}[tbh]
\begin{center}

\begin{tikzpicture}

\coordinate[label=below:{a)}] (A) at ( -1.5, 0.4);
\draw ( .3,0.)--( 1.7, 0.);
\draw ( 1., -.1)--( 1., .1);
\coordinate[label=below:{$(1)=\{2\}$}] (A) at ( 1., -0.2);
\coordinate[label=above:\tiny{$    1$}] (A) at (0.3, 0.);
\fill ( 0.3, 0.) circle (0.075);
\coordinate[label=above:\tiny{$    1$}] (A) at (1.7, 0.);
\fill ( 1.7, 0.) circle (0.075);

\coordinate[label=below:{b)}] (A) at ( -1.5, -2.1);
\draw ( -0.3,-2.5)--( 2.4, -2.5);
\coordinate[label=above:\tiny{$    1$}] (A) at (-0.3, -2.5);
\fill ( -0.3, -2.5) circle (0.075);
\coordinate[label=above:\tiny{$    1$}] (A) at (1.0, -2.5);
\fill ( 1, -2.5) circle (0.075);
\coordinate[label=above:\tiny{$    1$}] (A) at (2.4, -2.5);
\fill ( 2.4, -2.5) circle (0.075);
\coordinate[label=below:{$(2)=\{ 3\}$}] (A) at ( 1., -2.7);

\draw ( 5.5,0.)--( 9.7, 0.);
\draw ( 7.6, -.1)--( 7.6, .1);
\coordinate[label=above:\tiny{$    1$}] (A) at (5.5, 0.);
\fill ( 5.5, 0.) circle (0.075);
\coordinate[label=above:\tiny{$    1$}] (A) at (6.9, 0.);
\fill ( 6.9, 0.) circle (0.075);
\coordinate[label=above:\tiny{$    1$}] (A) at (8.3, 0.);
\fill ( 8.3, 0.) circle (0.075);
\coordinate[label=above:\tiny{$    1$}] (A) at (9.7, 0.);
\fill ( 9.7, 0.) circle (0.075);
\coordinate[label=below:{$(3)=\{ 4\}$}] (A) at ( 7.6, -0.2);

\draw ( 4.7, -2.5)--( 10.3, -2.5);
\coordinate[label=above:\tiny{$    1$}] (A) at (7.6, -2.5);
\fill ( 7.6, -2.5) circle (0.075);
\coordinate[label=above:\tiny{$    1$}] (A) at (6.2, -2.5);
\fill ( 6.2, -2.5) circle (0.075);
\coordinate[label=above:\tiny{$    1$}] (A) at (9.0, -2.5);
\fill ( 9.0, -2.5) circle (0.075);
\coordinate[label=above:\tiny{$    1$}] (A) at (4.7, -2.5);
\fill ( 4.7, -2.5) circle (0.075);
\coordinate[label=above:\tiny{$    1$}] (A) at (10.3, -2.5);
\fill ( 10.3, -2.5) circle (0.075);
\coordinate[label=below:{$(4)=\{ 5\}$}] (A) at ( 7.6, -2.7);

\coordinate[label=below:{c)}] (A) at ( -2., -5.6);
\draw ( -1.4,-6.)--( 9.8, -6.);
\fill [green] ( -1.4, -6.) circle (0.075);
\coordinate[label=below:\tiny{$    8$}] (A) at (-1.4, -6.);
\fill ( -0.7, -6.) circle (0.075);
\coordinate[label=below:\tiny{$    7$}] (A) at (-0.7, -6.);
\fill [green] ( 0.0, -6.) circle (0.075);
\coordinate[label=below:\tiny{$    6$}] (A) at (0.0, -6.);
\fill ( 0.7, -6.) circle (0.075);
\coordinate[label=below:\tiny{$    5$}] (A) at (0.7, -6.);
\fill [green] ( 1.4, -6.) circle (0.075);
\coordinate[label=below:\tiny{$    4$}] (A) at (1.4, -6.);
\fill ( 2.1, -6.) circle (0.075);
\coordinate[label=below:\tiny{$    3$}] (A) at (2.1, -6.);
\fill [green] ( 2.8, -6.) circle (0.075);
\coordinate[label=below:\tiny{$    2$}] (A) at (2.8, -6.);
\fill ( 3.5, -6.) circle (0.075);
\coordinate[label=below:\tiny{$    1$}] (A) at (3.5, -6.);
\draw ( 4.2, -6.1)--( 4.2, -5.9);
\fill ( 4.9, -6.) circle (0.075);
\coordinate[label=below:\tiny{$    1$}] (A) at (4.9, -6.);
\fill [green] ( 5.6, -6.) circle (0.075);
\coordinate[label=below:\tiny{$    2$}] (A) at (5.6, -6.);
\fill ( 6.3, -6.) circle (0.075);
\coordinate[label=below:\tiny{$    3$}] (A) at (6.3, -6.);
\fill [green] ( 7.0, -6.) circle (0.075);
\coordinate[label=below:\tiny{$    4$}] (A) at (7.0, -6.);
\fill ( 7.7, -6.) circle (0.075);
\coordinate[label=below:\tiny{$    5$}] (A) at (7.7, -6.);
\fill [green] ( 8.4, -6.) circle (0.075);
\coordinate[label=below:\tiny{$    6$}] (A) at (8.4, -6.);
\fill ( 9.1, -6.) circle (0.075);
\coordinate[label=below:\tiny{$    7$}] (A) at (9.1, -6.);
\fill [green] ( 9.8, -6.) circle (0.075);
\coordinate[label=below:\tiny{$    8$}] (A) at (9.8, -6.);

\coordinate[label=below:{$-$}] (A) at ( 1.4, -6.35);
\coordinate[label=below:{$+$}] (A) at ( 7.0, -6.35);

\coordinate[label=below:{$-$}] (A) at ( 1.4, -9.35);
\coordinate[label=below:{$+$}] (A) at ( 7.0, -9.35);

\coordinate[label=below:{d)}] (A) at ( -2., -8.6);
\draw ( -1.4,-9.)--( 9.8, -9.);
\fill [green] ( -1.4, -9.) circle (0.075);
\coordinate[label=below:\tiny{$    8$}] (A) at (-1.4, -9.);
\fill ( -0.7, -9.) circle (0.075);
\coordinate[label=below:\tiny{$    7$}] (A) at (-0.7, -9.);
\fill [green] ( 0.0, -9.) circle (0.075);
\coordinate[label=below:\tiny{$    6$}] (A) at (0.0, -9.);
\fill ( 0.7, -9.) circle (0.075);
\coordinate[label=below:\tiny{$    5$}] (A) at (0.7, -9.);
\fill [green] ( 1.4, -9.) circle (0.075);
\coordinate[label=below:\tiny{$    4$}] (A) at (1.4, -9.);
\fill ( 2.1, -9.) circle (0.075);
\coordinate[label=below:\tiny{$    3$}] (A) at (2.1, -9.);
\fill ( 2.8, -9.) circle (0.075);
\coordinate[label=below:\tiny{$    2$}] (A) at (2.8, -9.);
\fill ( 3.5, -9.) circle (0.075);
\coordinate[label=below:\tiny{$    1$}] (A) at (3.5, -9.);
\draw ( 4.2, -9.1)--( 4.2, -8.9);
\fill ( 4.9, -9.) circle (0.075);
\coordinate[label=below:\tiny{$    1$}] (A) at (4.9, -9.);
\fill ( 5.6, -9.) circle (0.075);
\coordinate[label=below:\tiny{$    2$}] (A) at (5.6, -9.);
\fill ( 6.3, -9.) circle (0.075);
\coordinate[label=below:\tiny{$    3$}] (A) at (6.3, -9.);
\fill ( 7.0, -9.) circle (0.075);
\coordinate[label=below:\tiny{$    4$}] (A) at (7.0, -9.);
\fill ( 7.7, -9.) circle (0.075);
\coordinate[label=below:\tiny{$    5$}] (A) at (7.7, -9.);
\fill ( 8.4, -9.) circle (0.075);
\coordinate[label=below:\tiny{$    6$}] (A) at (8.4, -9.);
\fill ( 9.1, -9.) circle (0.075);
\coordinate[label=below:\tiny{$    7$}] (A) at (9.1, -9.);
\fill [green] ( 9.8, -9.) circle (0.075);
\coordinate[label=below:\tiny{$    8$}] (A) at (9.8, -9.);
\draw [red]( 2.8, -9.)--( 8.4, -9);
\fill [red] ( 2.8, -9.) circle (0.075);
\coordinate[label=above:\tiny\color{red} {$    1$}] (A) at ( 2.8, -9.);
\fill [red] ( 4.2, -9.) circle (0.075);
\coordinate[label=above:\tiny\color{red} {$    1$}] (A) at ( 4.2, -9.);
\fill [red] ( 5.6, -9.) circle (0.075);
\coordinate[label=above:\tiny\color{red} {$    1$}] (A) at ( 5.6, -9.);
\fill [red] ( 7.0, -9.) circle (0.075);
\coordinate[label=above:\tiny\color{red} {$    1$}] (A) at ( 7.0, -9.);
\fill [red] ( 8.4, -9.) circle (0.075);
\coordinate[label=above:\tiny\color{red} {$    1$}] (A) at ( 8.4, -9.);

\coordinate[label=below:{$\{5\} \times \{2\}=\{6\}+\{4\}+\{2\}-\{2\}=\{6\}+\{4\}$} ] (A) at (  4.2, -10.);

\end{tikzpicture}

\end{center}
\caption{[Color online] \textit{Illustration of the Antoine-Speiser method for 
graphical tensor product reductions in $su(2)$. a) Weight diagrams of the two 
smallest half-integer spin representations $\{2\}$ and $\{4\}$, corresponding 
to $S = \frac{1}{2}$ and $S = \frac{3}{2}$, respectively. The superscripts 1
indicate that all states are non-degenerate. b) Weight diagrams of 
the integer spin representations $\{3\}$ and $\{5\}$, corresponding to 
$S = 1$ and $S = 2$. c) Landscape of $su(2)$ representations with a
positive sector on the right and a negative sector on the left. The integer
(black dots) and half-integer spin (green dots) representations are associated 
with the odd and even sublattice, respectively. d) The representation $\{5\}$ 
is centered at the position of $\{2\}$ in the landscape, resulting in the 
tensor product reduction $\{5\} \times \{2\} = \{6\} + \{4\} + \{2\} - \{2\} = 
\{6\} + \{4\}$.}}
\label{su2}
\end{figure}

\newpage

\newpage

\newpage

\newpage

\begin{figure}[tbp]
\begin{center}

%

\end{center}
\caption{[Color online] \textit{Landscape of irreducible $so(5) = sp(2)$ 
representations. The representations of trivial (black dots) and non-trivial 
(green dots) duality are associated with the odd and even sublattice of the 
square lattice, respectively. The position of a representation is characterized 
by the Cartesian coordinates $(x,y) = (2,1) + p \vec e_p + q \vec e_q$ with 
$\vec e_p = (1,0)$ and $\vec e_q = (1,1)$. The dimension $|D(p,q)|$ of 
eq.(\ref{Dsp2}) is listed below each point. The landscape is divided into
eight sectors of alternating signs $\pm$ (determined by the sign of $D(p,q)$),
each with a 45 degrees opening angle. The sectors are separated from each other 
by straight lines whose points are not associated with any irreducible 
representation.}}
\label{sp2landscape}
\end{figure}

\newpage

\newpage

\begin{figure}[tbp]
\begin{center}



\end{center}
\caption{[Color online] \textit{Tensor product reduction of the $so(5) = sp(2)$
representations $\{40\}$ and $\{16\}$. The weight diagram of $\{40\}$ is 
centered at the position of $\{16\}$ in the landscape. The degeneracy of a 
state in the weight diagram determines the multiplicity with which the
underlying representation in the landscape contributes to the tensor product
reduction. The sign $\pm$ of the contribution depends on which sector the
representation is positioned in.}}
\label{Speisersp2}
\end{figure}

\newpage

\newpage

\newpage

\newpage

\newpage

\begin{figure}[tbp]
\begin{center}

%

\end{center}
\caption{[Color online] \textit{Landscape of irreducible $su(3)$ 
representations. The representations of different triality are associated with
three distinct triangular sublattices shown in different colors. The position 
of a representation is characterized by the Cartesian coordinates 
$(x,y) = (\sqrt{3},0) + p \vec e_p + q \vec e_q$ with 
$\vec e_p = \frac{1}{2}(\sqrt{3},1)$ and 
$\vec e_q = \frac{1}{2}(\sqrt{3},-1)$. The dimension $|D(p,q)|$ of 
eq.(\ref{Dsu3}) is listed below each point. The landscape is divided into
six sectors of alternating signs $\pm$ (determined by the sign of $D(p,q)$),
each with a 60 degrees opening angle.}}
\label{su3landscape}
\end{figure}

\newpage

\newpage

\newpage

\newpage

\newpage

\begin{figure}[tbh]
\begin{center}

%

\end{center}
\caption{\it Landscape of irreducible $g(2)$ 
representations. The position of a representation is characterized by the
Cartesian coordinates 
$(x,y) = \frac{1}{2}(3 \sqrt{3},1) + p \vec e_p + q \vec e_q$ with 
$\vec e_p = (\sqrt{3},0)$ and $\vec e_q = \frac{1}{2}(\sqrt{3},1)$. The 
dimension $|D(p,q)|$ of eq.(\ref{Dg2}) is listed below each point. The 
landscape is divided into twelve sectors of alternating signs $\pm$ (determined
by the sign of $D(p,q)$), each with a 30 degrees opening angle.}
\label{g2landscape}
\end{figure}

\newpage

\end{document}